\documentclass[vecphys]{svmult}

\usepackage{makeidx}         % allows index generation
\usepackage{graphicx}        % standard LaTeX graphics tool
\usepackage{multicol}        % used for the two-column index
\usepackage[bottom]{footmisc}% places footnotes at page bottom

\usepackage{psfig}

\newcommand{\be}{\begin{equation}}
\newcommand{\ee}{\end{equation}}
\newcommand{\ba}{\begin{eqnarray}}
\newcommand{\ea}{\end{eqnarray}}
\newcommand{\brr}{\begin{array}}
\newcommand{\err}{\end{array}}
\newcommand{\bc}{\begin{center}}
\newcommand{\ec}{\end{center}}
\newcommand{\hm}{\,h^{-1}{\rm Mpc}}

\newcommand{\msun}{\,h^{-1}M_\odot}

\newcommand{\lb}{{\left<\right.}}
\newcommand{\rb}{{\left.\right>}}
\newcommand{\lum}{\,{\rm erg\,s^{-1}}}
\newcommand{\vel}{\,{\rm km\,s^{-1}}}

\newcommand{\bx}{\rm{\bf x}}
\newcommand{\by}{\rm{\bf y}}
\newcommand{\bk}{\rm{\bf k}}
\newcommand{\bv}{\rm{\bf v}}
\newcommand{\bu}{\rm{\bf u}}
\newcommand{\br}{\rm{\bf r}}
\newcommand{\tv}{t_{\rm vir}}

\newcommand{\mincir}{\raise
  -2.truept\hbox{\rlap{\hbox{$\sim$}}\raise5.truept \hbox{$<$}\ }}
\newcommand{\magcir}{\raise
  -2.truept\hbox{\rlap{\hbox{$\sim$}}\raise5.truept \hbox{$>$}\ }}
\newcommand{\siml}{\raise
  -2.truept\hbox{\rlap{\hbox{$\sim$}}\raise5.truept \hbox{$<$}\ }}
\newcommand{\simg}{\raise
  -2.truept\hbox{\rlap{\hbox{$\sim$}}\raise5.truept \hbox{$>$}\ }}

\newcommand{\aj}{AJ}
\newcommand{\apj}{ApJ}
\newcommand{\apjl}{ApJ}
\newcommand{\apjs}{ApJS}
\newcommand{\aap}{A\&A}
\newcommand{\mnras}{MNRAS}
\newcommand{\nat}{Nature}
\newcommand{\araa}{ARAA}
\newcommand{\prd}{Phys. Rev. D}
\newcommand{\physrep}{Phys. Rep.}

\makeindex             % used for the subject index
\begin{document}

\title*{Cosmology with clusters of galaxies}
\author{Stefano Borgani}
\institute{Department of Astronomy, University of Trieste, via Tiepolo
  11, I-34131 Trieste (Italy)
\texttt{borgani@ts.astro.it}}
\maketitle

%Your text goes here. Separate text sections with the standard \LaTeX\
%sectioning commands.

\section{Introduction}
\label{intro}
Clusters of galaxies occupy a special place in the hierarchy of cosmic
structures. They arise from the collapse of initial perturbations
having a typical comoving scale of about 10$\hm$\footnote{Here $h$ is
the Hubble constant in units of 100$\vel$ Mpc$^{-1}$}. According to
the standard model of cosmic structure formation, the Universe is
dominated by gravitational dynamics in the linear or weakly
non--linear regime and on scales larger than this. In this case, the
description of cosmic structure formation is relatively simple since
gas dynamical effects are thought to play a minor role, while the
dominating gravitational dynamics still preserves memory of initial
conditions. On smaller scales, instead, the complex astrophysical
processes, related to galaxy formation and evolution, become
relevant. Gas cooling, star formation, feedback from supernovae (SN)
and active galactic nuclei (AGN) significantly change the evolution of
cosmic baryons and, therefore, the observational properties of the
structures. Since clusters of galaxies mark the transition between
these two regimes, they have been studied for decades both as
cosmological tools and as astrophysical laboratories.

In this Chapter I concentrate on the role that clusters play in
cosmology. I will highlight that, in order for them to be calibrated
as cosmological tools, one needs to understand in detail the
astrophysical processes which determine their observational
characteristics, i.e. the properties of the cluster galaxy population
and those of the diffuse intra--cluster medium (ICM).

Constraints of cosmological parameters using galaxy clusters have been
placed so far by applying a variety of methods. 
For example:
\begin{enumerate}
\item[1.] The mass function of nearby galaxy clusters provides
  constraints on the amplitude of the power spectrum at the
  cluster scale (e.g., \cite{2002ARA&A..40..539R,2005RvMP...77..207V} 
  and references  therein). At the same time,
  its evolution provides constraints on the linear growth rate of
  density perturbations, which translate into dynamical constraints
  on the matter and Dark Energy (DE) density parameters.
\item[2.] The clustering properties (i.e., correlation function and
  power spectrum) of the large--scale distribution of galaxy clusters
  provide direct information on the shape and amplitude of the
  underlying DM distribution power spectrum. Furthermore, the
  evolution of these clustering properties is again sensitive to the
  value of the density parameters through the linear growth rate of
  perturbations (e.g., \cite{2001Natur.409...39B,2001MNRAS.327..422M}
  and references therein).
\item[3.] The mass-to-light ratio in the optical band can be used to
  estimate the matter density parameter, $\Omega_m$, once the mean
  luminosity density of the Universe is known and under the assumption
  that mass traces light with the same efficiency both inside and
  outside clusters (see
  \cite{2000ApJ...541....1B,2000ApJ...530...62G,1996ApJ...462...32C},
  as examples of the application of this method).
\item[4.] The baryon fraction in nearby clusters provides constraints
  on the matter density parameter, once the cosmic baryon density
  parameter is known, under the assumption that clusters are fair
  containers of baryons (e.g.,
  \cite{1991MNRAS.253P..29F,1993Natur.366..429W}). Furthermore, the
  baryon fraction of distant clusters provide a geometrical constraint
  on the DE content and equation of state, under the additional
  assumption that the baryon fraction within clusters does not evolve
  (e.g., \cite{2002MNRAS.334L..11A,2003A&A...398..879E}).
\end{enumerate}

An extensive presentation of all these methods would probably require
a dedicated book. For this reason, in this contribution I will mostly
concentrate on the method based on the evolution of the cluster mass
function. A substantial part of my Lecture will concentrate on the
different methods that have been applied so far to weight galaxy
clusters. Since all the above cosmological applications rely on
precise measurements of cluster masses, this part of my contribution
will be of general relevance for cluster cosmology.

Also, since most of the cosmological applications of galaxy clusters
have been based so far on X--ray surveys, my discussion will be
definitely X--ray biased, although I will discuss in some detail
methods based on optical observations and what present and future
optical surveys are expected to provide. I refer to the Lectures by
R. Gal and O. Lopez--Cruz in this volume for more details regarding
the properties of galaxy clusters in the optical band. Also, I will
refer to the Lectures by M. Birkinshaw for cosmological studies of
clusters based on the Sunyaev--Zeldovich (SZ) effect, to the Lectures
by J.--P. Kneib for cluster studies and mass measurement through
gravitational lensing, and to the Lectures by C. Jones and by
C. Sarazin for more details about the cosmological application of the
baryon fraction method.

The structure of this Chapter will be as follows. I provide in Section
2 a short introduction to the basics of cosmic structure formation. I
will shortly review the linear theory for the evolution of density
perturbations and the spherical collapse model. In Section 3 I will
describe the Press--Schechter (PS) formalism to derive the
cosmological mass function. I will then introduce extensions of the PS
approach and present the most recent calibrations of the mass function
from N--body simulations. In Section 4 I will review the methods to
build samples of galaxy clusters, based on optical and X--ray
observations, while I will only briefly discuss the SZ methodology for
cluster surveys. Section 5 is devoted to the discussion of different
methods to derive cluster masses and to review the results of the
application of these methods. In Section 6 I will describe the
cosmological constraints, which have been obtained so far by tracing
the cluster mass function with a variety of methods: distribution of
velocity dispersions, X--ray temperature and luminosity functions, and
gas mass function. In this Section I will also critically discuss the
reasons for the different, sometimes discrepant, results that have
been obtained in the literature and I will highlight the relevance of
properly including the analysis of the cluster mass function all the
statistical and systematic uncertainties in the relation between mass
and observables. Finally, I will describe in Section 7 the future
perspectives for cosmology with galaxy clusters and which are the
challenges for clusters to keep playing an important role in the era
of precision cosmology.

\section{A concise handbook of cosmic structure formation}
In this section I will briefly review the basic concepts of cosmic
structure formation, which are relevant for the study of galaxy
clusters as tools for precision cosmology through the evolution of
their mass function. A complete treatment of models of structure
formation can be found in classical cosmology textbooks (e.g.,
\cite{1993ppc..book.....P,1999coph.book.....P,2002coec.book.....C}).

\subsection{The statistics of cosmic density fields}
Let $\rho(\bx)$ be the matter density field, which is a continuous
function of the position vector $\bx$, $\bar\rho=\lb \rho\rb$ its
average value computed over a sufficiently large (representative)
volume of the Universe and
\be \delta(\bx)={\rho(\bx)-\bar\rho \over
  \bar\rho}
\label{eq:del}
\ee
the corresponding relative density contrast. By definition, it is
$\bar\delta=0$ and $\delta(\bx)\ge -1$. If the density field
is traced by a discrete distribution of points (i.e., galaxies or galaxy
clusters) all having the same weight (mass), then $\rho(\bx)=\sum_i
\delta_D(\bx-\bx_i)$, where $\delta_D(\bx)$ is the Dirac
delta--function. The Fourier representation of the density contrast is
given by 
\be
\tilde\delta(\bk)={1\over (2\pi)^{3/2}}\int d\bx \,\delta(\bx)e^{i\bk\cdot
  \bx}\,, 
\ee
with the corresponding dual relation for the inverse Fourier
transform. 

The 2--point correlation function for the density contrast is defined
as 
\be
\xi(r)=\lb \delta(\bx_1)\delta(\bx_2)\rb \,,
\ee
which only depends on the modulus of the separation vector, $r=|\bx_1
-\bx_2|$, under the assumption of statistical isotropy of the density
field. Therefore, it can be shown that the power spectrum of the
density fluctuations is the Fourier transform of the correlation
function, so that 
\be
P(k)=\lb |\tilde\delta(\bk)|^2\rb ={1\over 2\pi^2}\int dr \,r^2 \xi(r) {\sin
kr \over kr}\,,
\ee
which, again, depends only on the modulus of the wave-vector $\bk$. 

In case we are interested in the study of a class of observable
structures of mass $M$, which arise from the collapse of initial
perturbations having size $R\propto (M/\bar\rho)^{1/3}$, then it is
common to introduce the smoothed density field, which is defined as
\be
\delta_R(\bx)= \delta_M(\bx)=\int \delta(\by)W_R(|\bx-\by|)\,d\by\,.
\label{eq:delr}
\ee
As such, it is given by the convolution of the density fluctuation field
with a window function, which filters out the fluctuation modes having
wavelength $\mincir R$. Eq.(\ref{eq:delr}) allows us to introduce
the variance of the fluctuation field computed at the scale R, 
defined as
\be 
\sigma^2_R=\sigma^2_M = \lb \delta_R^2\rb = {1\over
  2\pi^2}\int dk \,k^2P(k)\,\tilde W_R^2(k)\,,
\label{eq:var}
\ee
where $\tilde W_R(k)$ is the Fourier transform of the window function. 

The shape of the window function defines the exact relation between
mass and smoothing scale. For instance, for the top--hat window it is 
\be
\tilde W_R(k)={3[\sin(kR)-kR\cos(kR)]\over (kR)^3} 
\label{eq:thw}
\ee 
while the Gaussian window gives
\be
\tilde W_R(k)=\exp\left(-{(kR)^2\over 2} \right),.
\label{eq:gauw}
\ee 
The corresponding relations between mass scale and smoothing scale are
$M=(4\pi/3)R^3\bar\rho$ and $M=(2\pi R^2)^{3/2}\bar\rho$ for the
top--hat and Gaussian filters, respectively. 

The shape of the power spectrum is (essentially) fixed once the matter
density parameter, $\Omega_m$, that associated to the baryonic
component, $\Omega_{\rm bar}$, and the Hubble parameter, $H_0$, are
specified (e.g., \cite{1999ApJ...511....5E}). However, its
normalization can only be fixed through a comparison with
observational data of the large--scale structure of the Universe or of
the anisotropies of the Cosmic Microwave Background (CMB). A common
way of parametrizing this normalization is through the quantity
$\sigma_8$, which is defined as the variance, computed for a top--hat
window having comoving radius $R=8\hm$ (given in eq. \ref{eq:var}).
The historical reason for this choice of the normalization scale is
that the variance of the galaxy number counts, within the first
redshift surveys, was observed to be about unity inside spheres of
that radius (e.g., \cite{1983ApJ...267..465D}). In this way, the value
of $\sigma_8$ for a given cosmology directly provides a measure of the
biasing parameter relating the galaxy and mass distribution, expected
for that model. Furthermore, a top--hat sphere of $8\hm$ radius
contains a mass $M\simeq 5.9\times 10^{14}\Omega_m\msun$, which is the
typical mass of a moderately rich galaxy cluster. Therefore, as we
shall see in Section \ref{s:mf}, the mass function of galaxy clusters
provides a direct measure of $\sigma_8$.

\subsection{The linear evolution of density perturbations}
Let us assume that the matter content of the Universe is dominated by
a pressurless and self--gravitating fluid. This approximation holds if
we are dealing with the evolution of the perturbations in the dark
matter (DM) component or in case we are dealing with structures whose
size is much larger than the typical Jeans scale--length of
baryons. Let us also define $\bx$ to be the comoving coordinate and
$\br = a(t)\bx$ the proper coordinate, $a(t)$ being the cosmic
expansion factor. Furthermore, if $\bv =\dot{\br}$ is the physical
velocity, then $\bv =\dot a\bx+\bu$, where the first term describes
the Hubble flow, while the second term, $\bu=a(t)\dot{\bx}$, gives the
peculiar velocity of a fluid element which moves in an expanding
background.

In this case the equations that regulate the Newtonian description of
the evolution of density perturbations are the continuity equation:
\be
{\partial \delta\over \partial t}+\nabla \cdot [(1+\delta)\bu]=0\,,
\label{eq:cont}
\ee
which gives the mass conservation, the Euler equation
\be
{\partial \bu\over \partial t} + 2H(t)\bu + (\bu\cdot\nabla)\bu =
-{\nabla \phi\over a^2}\,,
\label{eq:eul}
\ee
which gives the relation between the acceleration of the fluid element
and the gravitational force, and the Poisson equation
\be
\nabla^2\phi = 4\pi G\bar\rho a^2\delta
\label{eq:pois}
\ee
which specifies the Newtonian nature of the gravitational force. In
the above equations, $\nabla$ is the gradient computed with
respect to the comoving coordinate $\bx$, $\phi(\bx)$ describes the
fluctuations of the gravitational potential and $H(t)=\dot a/a$ is the
Hubble parameter at the time $t$. Its time--dependence is given by
$H(t)=E(t)H_0$, where 
\be
E(z)=[(1+z)^3\Omega_m+(1+z)^2(1-\Omega_m-\Omega_{DE})+(1+z)^{3(1+w)}
\Omega_{DE}]^{1/2} 
\label{eq:ez}
\ee
is related to the density parameter contributed by non--relativistic
matter, $\Omega_m$, and by Dark Energy (DE), $\Omega_{DE}$, with
equation of state $p=w\rho c^2$ (if the DE term is
provided by cosmological constant then $w=-1$).

In the case of small perturbations, these equations can be linearized
by neglecting all the terms which are of second order in the fields
$\delta$ and $\bu$. In this case, after further differentiating
eq.(\ref{eq:cont}) with respect to time, using the Euler equation to
eliminate the term $\partial \bu/\partial t$, and using the Poisson
equation to eliminate $\nabla^2\phi$, one ends up with:
\be
{\partial^2 \delta\over \partial t^2}+2H(t){\partial
  \delta\over \partial t}-4\pi G\bar\rho\delta=0\,.
\label{eq:linear}
\ee
This equation describes the Jeans instability of a pressurless fluid,
with the additional ``Hubble drag'' term $2H(t) {\partial \delta /
\partial t}$, which describes the counter--action of the expanding
background on the perturbation growth. Its effect is to prevent the
exponential growth of the gravitational instability taking place in a
non--expanding background \cite{1987gady.book.....B}.  The solution
of the above equation can be casted in the form:
\be
\delta(\bx,t) = \delta_+(\bx,t_i)D_+(t)+\delta_-(\bx,t_i)D_-(t)\,,
\label{eq.linsol}
\ee
where $D_+$ and $D_-$ describes the growing and decaying modes of the
density perturbation, respectively. In the case of an
Einstein--de-Sitter (EdS) Universe ($\Omega_m=1$, $\Omega_{DE}=0$), it
is $H(t)=2/(3t)$, so that $D_+(t)=(t/t_i)^{2/3}$ and $D_-(t)=
(t/t_i)^{-1}$. The fact that $D_+(t)\propto a(t)$ for an EdS Universe
should not be surprising. Indeed, the dynamical time--scale for the
collapse of a perturbation of uniform density $\rho$ is $t_{\rm
dyn}\propto (G\rho)^{-1//2}$, while the expansion time scale for the
EdS model is $t_{\rm exp}\propto (G\bar\rho)^{-1//2}$, where
$\bar\rho$ is the mean cosmic density. Since for a linear (small)
perturbation it is $\rho \simeq \bar\rho$, then $t_{\rm dyn}\sim
t_{\rm exp}$, thus showing that the cosmic expansion and the
perturbation evolution take place at the same pace. This argument also
leads to understanding the behaviour for a $\Omega_m<1$ model. In this
case, the expansion time scale becomes shorter than the above one at
the redshift at which the Universe recognizes that $\Omega_m<1$. This
happens at $1+z\simeq \Omega_m^{-1/3}$ or at $1+z\simeq \Omega_m^{-1}$
in the presence or absence of a cosmological constant term,
respectively. Therefore, after this redshift, cosmic expansion takes
place at a quicker pace than gravitational instability, with the
result of freezing the perturbation growth.

\begin{figure}
\centering
\includegraphics[height=8.5cm]{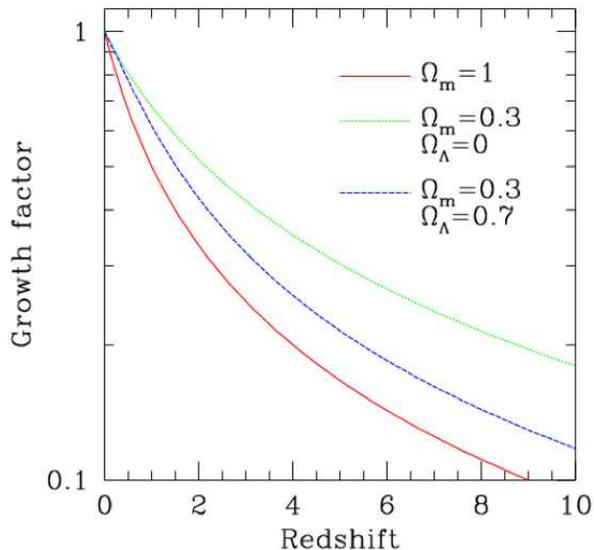}
\caption{The redshift dependence of the linear growth mode of
  perturbations for a flat model with $\Omega_m=1$ (solid curve), 
  for a flat $\Omega_m=0.3$ model with a cosmological constant (dashed
  curve) and for an $\Omega_m=0.3$ open model with vanishing
  cosmological constant (dotted curve). }
\label{fi:grlin}
\end{figure}

\begin{figure}
\centerline{
\psfig{file=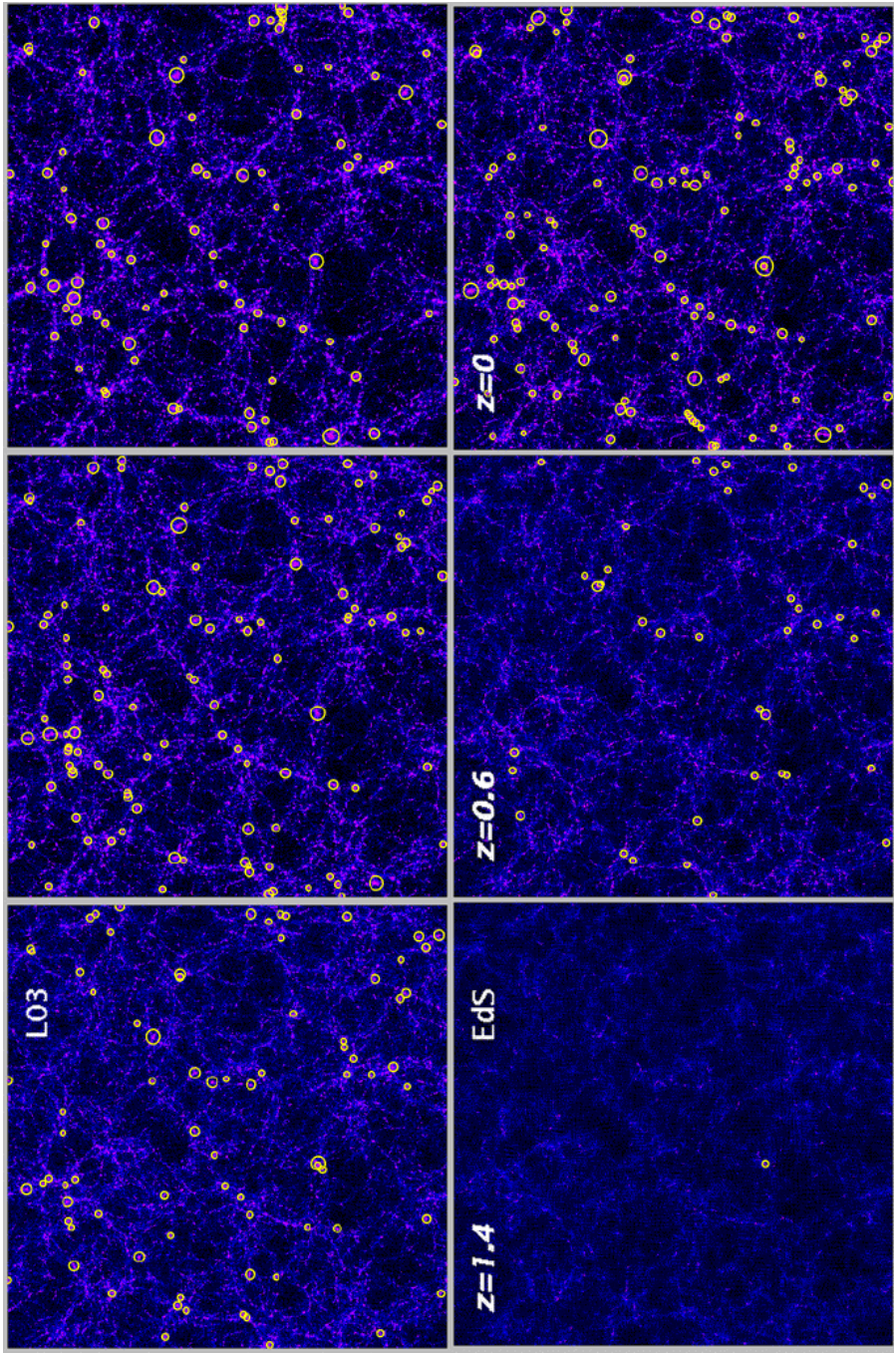,height=18cm} 
}
\caption{The evolution of the cluster population from N--body
simulations in two different cosmologies \cite{2001Natur.409...39B}.
Left panels describe a flat, low--density model with $\Omega_m=0.3$
and $\Omega_\Lambda=0.7$ (L03); right panels are for an
Einstein--de-Sitter model (EdS) with $\Omega_m=1$. Superimposed on the
dark matter distribution, the yellow circles mark the positions of
galaxy clusters with virial temperature $T>3$ keV, the size of the
circles is proportional to temperature. Model parameters have been
chosen to yield a comparable space density of clusters at the present
time.  Each snapshot is $250h^{-1}$ Mpc across and $75h^{-1}$ Mpc
thick (comoving with the cosmic expansion). }
\label{fi:natfig}
\end{figure}

The exact expression for the growing model of perturbations is given
by 
\be
D_+(z)\,=\,{5\over 2}\,\Omega_m E(z)\,\int_z^\infty {1+z'\over
E(z')^3}\, dz'
\label{eq:grw}
\ee 
(e.g., \cite{1993ppc..book.....P}). I show in Figure \ref{fi:grlin}
the redshift dependence of the linear growth factor for an Eds model
and for two models with $\Omega_m=0.3$ both with and without a
cosmological constant term to restore spatial flatness. Quite
apparently, the EdS has the faster evolution, while the slowing down
of the perturbation growth is more apparent for the open low--density
model, the presence of cosmological constant providing an intermediate
degree of evolution. A more pictorial view is provided in Figure
\ref{fi:natfig}, where we show the dark matter density fields for two
different cosmologies and at different epochs, as obtained from N--body
simulations. The two models, an EdS one and a flat low--density one
with $\Omega_m=0.3$, have been tuned so as to have a similar
appearance at $z=0$. This figure clearly shows that any observational
probe of the degree of evolution of density perturbations would
correspond to a sensitive probe of cosmological parameters. Such a
cosmological test is conceptually different to those provided by the
standard geometrical tests based on luminosity and angular--size
distances.

As we shall discuss in the following, clusters of galaxies provide
such a probe, since the evolution of their number density is
directly related to the growth rate of perturbations.

\subsection{The spherical top--hat collapse}
A spherical perturbation at constant density represents the only case
in which the evolution can be exactly computed. Although the
assumptions on which this model is based are quite restrictive,
nevertheless it serves as a very useful guideline to characterize the
process of evolution and formation of virialized DM halos. This
approach is based on treating the perturbation as a separate
Friedmann--Lemaitre--Robertson--Walker (FLRW) universe, with the
constraint of null velocity at the boundary of the perturbation. Here
we will sketch the derivation in the case of $\Omega_m=1$ (e.g.,
\cite{2002coec.book.....C}, while an extension of this derivation to
more general cosmologies can be found in 
\cite{1996MNRAS.282..263E} and \cite{1996ApJ...469..480K}, with useful
fitting functions provided in \cite{1998ApJ...495...80B}.

Assuming null velocities at an initial time $t_i$ provides the
relation $D_+(t_i)=(3/5)\delta(t_i)$, between the linear growth mode of
the perturbation and the initial overdensity. The initial density
parameter, which characterizes this separate Universe, is then
$\Omega_p(t_i)=\Omega(t_i)(1+\delta_i)$. Therefore, the condition for
the perturbation to recollapse will be $\Omega_p(t_i)>1$. If this
condition is satisfied, then we can derive the density within the
perturbation at the time $t_m$ of its maximum expansion (turn--around)
as
\be
\rho_p(t_m)=\rho_c(t_i)\Omega_p(t_i)\left[{\Omega_p(t_i)-1\over
    \Omega_p(t_i)}\right]^3\,.
\ee
The time $t_m$ is given by the solution of the Friedmann equations for
a closed Universe:
\be t_m={\pi\over 2H_i}{\Omega_p(t_i)\over
  \left[\Omega_p(t_i)-1\right]^{3/2}}=\left[{3\pi \over
    32G\rho_p(t_m)} \right]\,, \ee
where $H_i$ is the Hubble parameter within the perturbation. At the
same epoch $t_m$, the density of the general cosmic background is
$\rho(t_m)=(6\pi Gt_m^2)^{-1}$. Therefore, the exact value for the
perturbation overdensity at the turn--around is
\be
\delta_+(t_m)={\rho_p(t_m)\over \rho(t_m)}-1=\left( {3\pi \over
    4}\right)^2 -1 \simeq 4.6\,.
\label{eq:ta}
\ee
On the other hand, the linear--theory extrapolation to $t_m$ would
give
\be
\delta_+(t_m)=\delta_+(t_i)\left( {t_m\over
    t_i}\right)^{2/3}={3\over 5}\left( {3\pi \over
    4}\right)^{2/3}\simeq 1.07\,.
\label{eq:talin}
\ee
This demonstrates that the linear--theory extrapolation significantly
underestimates overdensities at the turn-around.

After reaching the maximum expansion, the perturbation then evolves by
detaching from the general Hubble expansion and then recollapses,
reaching virial equilibrium supported by the velocity dispersion of DM
particles. This happens at the virialization time $\tv$, at which the
perturbation meets by definition the virial condition $E=K+U=-K$,
being $E$, $K$ and $U$ the total, the kinetic and the potential
energy, respectively.

At the turn--around point, the perturbation has no kinetic energy, so
that the total energy is
\be
E_m=U=-{3\over 5}{GM^2\over R_m} \;\;,
\ee
where we have used the expression for the potential energy of a
uniform spherical density field of radius $R_m$ and total mass $M$. In
a similar manner, the total energy at the virialization is
\be
E_{\rm vir}={U\over 2}=-{1\over 2}{3\over 5}{GM^2\over R_{\rm vir}}\,.
\ee
Therefore, the condition of energy conservation in a dissipationless
collapse gives $R_m=2R_{\rm vir}$ for the relation between the radii
at turn-around and at virial equilibrium. This allows us to compute
the overdensity at $\tv$ as
\be
{\rho_p(\tv)\over \rho(\tv)}=\left( {\tv \over t_m}\right)^2 \left(
  {R_m\over R_{\rm vir}}\right)^3{\rho_p(t_m)\over \rho(t_m)}=2^2 2^3
\left( {3\pi \over 4}\right)^2=18\pi^2\simeq 178\,,
\label{eq:vir}
\ee
where we have accounted for both the compression of the perturbation
density, due to its shrinking, and of the dilution of the background
density as the Universe expands from $t_m$ to $\tv$.
Eq.(\ref{eq:vir}) shows why an overdensity of about 200 is usually
considered as typical for a DM halo which has reached the condition of
virial equilibrium. As for the extrapolation of linear--theory
prediction, it would have given
\be
\delta_+(\tv)=\left({\tv \over t_m} \right)^{2/3}\delta_+(t_m)\simeq
1.69\,.
\label{eq:virlin}
\ee
The above equation shows the derivation of another fundamental number
that will be used in what follows in order to characterize the mass
function of virialized halos. It gives the overdensity that a
perturbation in the initial density field must have for it to end up
in a virialized structure. While the above derivation holds for an EdS
Universe, it can be generalized to any generic cosmology. For
$\Omega_m<1$ the increased expansion rate of the Universe causes a
faster dilution of the cosmic density from $t_m$ to $\tv$ and, as a
consequence, a larger value of the overdensity at virialization.

In the following, we will indicate with $\Delta_{\rm vir}$ the
overdensity at virial equilibrium, computed with respect to the
background density, and with $\Delta_c$ the same quantity expressed in
units of the critical density $\rho_c$. As a reference, a flat
low--density model with $\Omega_m=0.3$ has $\Delta_c\simeq 100$ and
$\Delta_{\rm vir}\simeq 330$. Also, we will use in the following the
notation $R_N$ to indicate the radius of a halo encompassing an
average overdensity equal to $N\rho_c$, so that $M_N$ will denote the
halo mass contained within that radius. As we shall see in the
following, values often used in the literature are $N=200$, 500 and
2500.

\section{The mass function}
\label{s:mf}
The mass function (MF) at redshift $z$, $n(M,z)$, is defined as the
number density of virialized halos found at that redshift with mass in
the range $[M,M+dM]$. In this section I will derive the MF expression
following the approach originally devised by Press and Schechter
\cite{PR74.1} (PS hereafter). After commenting on the limitations of
this approach, I will discuss the accuracy with which improved
derivations of the MF reproduce the ``exact'' predictions from N--body
simulations.

\subsection{The Press--Schechter mass function}
The PS derivation of the MF is based on the assumption that the
fraction of matter ending up in objects of a given mass $M$ can be
found by looking at the portion of the initial (Lagrangian) density
field, smoothed on the mass--scale $M$, lying at an overdensity
exceeding a given critical threshold value, $\delta_c$. Under the
assumption of Gaussian perturbations, the probability for the
linearly--evolved smoothed field $\delta_M$ to exceed at redshift $z$
the critical density contrast $\delta_c$ reads
\be p_{>\delta_c}(M,z)={1\over \sqrt{2\pi}
  \sigma_M(z)}\int_{\delta_c}^\infty \exp\left(-{\delta_M^2\over
    2\sigma_M(z)^2} \right)\,d\delta_M={1\over 2}{\rm erfc}\left(
  {\delta_c\over \sqrt 2\sigma_M(z)}\right)\,,
\label{eq:frvol}
\ee
where ${\rm erfc}(x)$ is the complement error function and
$\sigma_M(z)=\delta_+(z)\sigma_M$ is the variance at the mass scale
$M$ linearly extrapolated at redshift $z$. Under the assumption of
spherical collapse, the critical overdensity $\delta_c$ is given by
the linear extrapolation of the overdensity at virial equilibrium, as
derived in the previous section. In this case, it will be
$\delta_c=\delta_c(z)$ with a weak dependence upon redshift and
cosmological parameters, with $\delta_c\simeq 1.69$ independent of $z$
only in the case of an EdS cosmology. By definition, the above
equation provides the fraction of unity volume, which ends up by
redshift $z$ in objects with mass above $M$. Therefore, the fraction
of Lagrangian volume in objects with mass in the range $[M,M+dM]$ is
\be 
dp_{>\delta_c}(M,z)=\left|{\partial
    p_{>\delta_c}(M,z)\over \partial M }\right| dM\,.
\label{eq:diffpr}
\ee
Since the probability of eq.(\ref{eq:frvol}) is a decreasing function
of mass, the absolute value is required in order to have a
positive--defined differential probability. Eq.(\ref{eq:diffpr}) shows
a fundamental limitation of the PS derivation of the MF. Indeed, we
expect that, as we take the limit of arbitrarily small limiting mass,
we should recover the whole mass content of the Universe. This is to
say that, in the hierarchical clustering picture, all the mass is
contained within halos of arbitrarily small mass. However, integrating
eq.(\ref{eq:diffpr}) over the whole mass range gives $\int_0^\infty
dp_{>\delta_c}(M,z)=1/2$. This implies that the PS derivation of the
mass function only accounts for half of the total mass at disposition.
The basic reason for this is that, in this derivation, we give zero
probability for a point with $\delta_M<\delta_c$, for a given
filtering mass scale $M$, to have $\delta_{M'}>\delta_c$ for some
larger filtering scale $M'>M$. This means that the PS approach
neglects the possibility for that point to end up in a collapsed halo
of larger mass. A more rigorous derivation of the mass function, which
is based on the excursion--set formalism
\cite{1991ApJ...379..440B}, correctly accounts for the missing factor
2, at least for the particular choice of a sharp--$k$ filter (i.e., a
top--hat window function in Fourier space).

Since eq.(\ref{eq:diffpr}) provides the fraction of volume in objects of
a given mass, the number density of such objects will be obtained
after dividing it by the volume, $V_M=M/\bar\rho$, occupied by each
object. Therefore, after accounting for the missing factor 2, the
expression for the mass function reads
\ba
{dn(M,z) \over dM} & = & {2\over V_M}
{\partial p_{>\delta_c}(M,z)\over \partial M} \nonumber \\
& = & \sqrt{2\over \pi}{\bar\rho \over M^2}{\delta_c\over
  \sigma_M(z)}\left| {d\log \sigma_M(z)\over d\log M}\right| \exp\left(-{\delta_c^2\over
    2\sigma_M(z)^2} \right)\,.
\label{eq:psmf}
\ea
This is the expression for the PS mass function. Although we will
present below a more accurate expressions for the MF, this equation
already demonstrates the reason for which the mass function of galaxy
clusters is a powerful probe of cosmological models. Cosmological
parameters enter in eq.(\ref{eq:psmf}) through the mass variance
$\sigma_M$, which depends on the power spectrum and on the
cosmological density parameters, through the linear perturbation
growth factor, and, to a lesser degree, through the critical density
contrast $\delta_c$. Taking this expression in the limit of massive
objects (i.e., rich galaxy clusters), the MF shape is dominated by the
exponential tail. This implies that the MF becomes exponentially
sensitive to the choice of the cosmological parameters. In other
words, a reliable observational determination of the MF of rich
clusters would allow us to place tight constraints on cosmological
parameters.

\subsection{Extensions of the PS approach and N--body tests}
Following \cite{JE01.1}, an alternative way of recasting the mass
function is  
\be
f(\sigma_M,z)={M\over \bar\rho}{dn(M,z)\over d\ln \sigma_M^{-1}}\,.
\ee
In this way, the PS expression is recovered by setting
\be
f(\sigma_M,z)=\sqrt{2\over \pi}{\delta_c\over
  \sigma_M}\exp\left( -{\delta_c^2\over 2\sigma_M^2}\right) 
\ee
Despite its subtle simplicity (e.g., \cite{1998FCPh...19..157M}),
the PS MF has served for more than a decade as a guide to constrain
cosmological parameters from the mass distribution of galaxy
clusters. Only with the advent of a new generation of N--body
simulations, which are able to cover a very large dynamical range,
have significant deviations of the PS expression from the exact
numerical description been noticed (e.g., 
\cite{1998MNRAS.301...81G,1999MNRAS.307..949G,JE01.1,2002ApJ...573....7E,2005Natur.435..629S,2005astro.ph..6395W}). Such
deviations have been usually interpreted in terms of corrections to
the PS approach.

Incorporating the effect of non--spherical collapse, 
the PS expression has been generalized \cite{SH99.1} to
\be 
f(\sigma_M,z)=\sqrt{2a\over \pi}C\left[ 1+\left({\sigma_M^2\over
a\delta_c^2}\right)^q\right]{\delta_c\over \sigma_M} 
\exp\left(-{a\delta_c^2\over 2\sigma_M^2} \right)\,.
\label{eq:stmf}
\ee
These authors also compared this expression with results from N--body
simulations, in which the mass of the clusters were estimated with a
spherical overdensity (SO) algorithm, by computing the mass within the
radius encompassing a mean overdensity equal to the virial one. As a
result, they found the best--fitting values $a=0.707$, $q=0.3$, with
the normalization constant $C=0.3222$ obtained from the normalization
requirement $\int_0^\infty f(\sigma_M)d\nu=1$ (note that the PS
expression is recovered for $a=1$, $q=0$ and $C=1/2$; see also
\cite{2002MNRAS.329...61S}).

\begin{figure}
\centerline{
\hbox{
\psfig{file=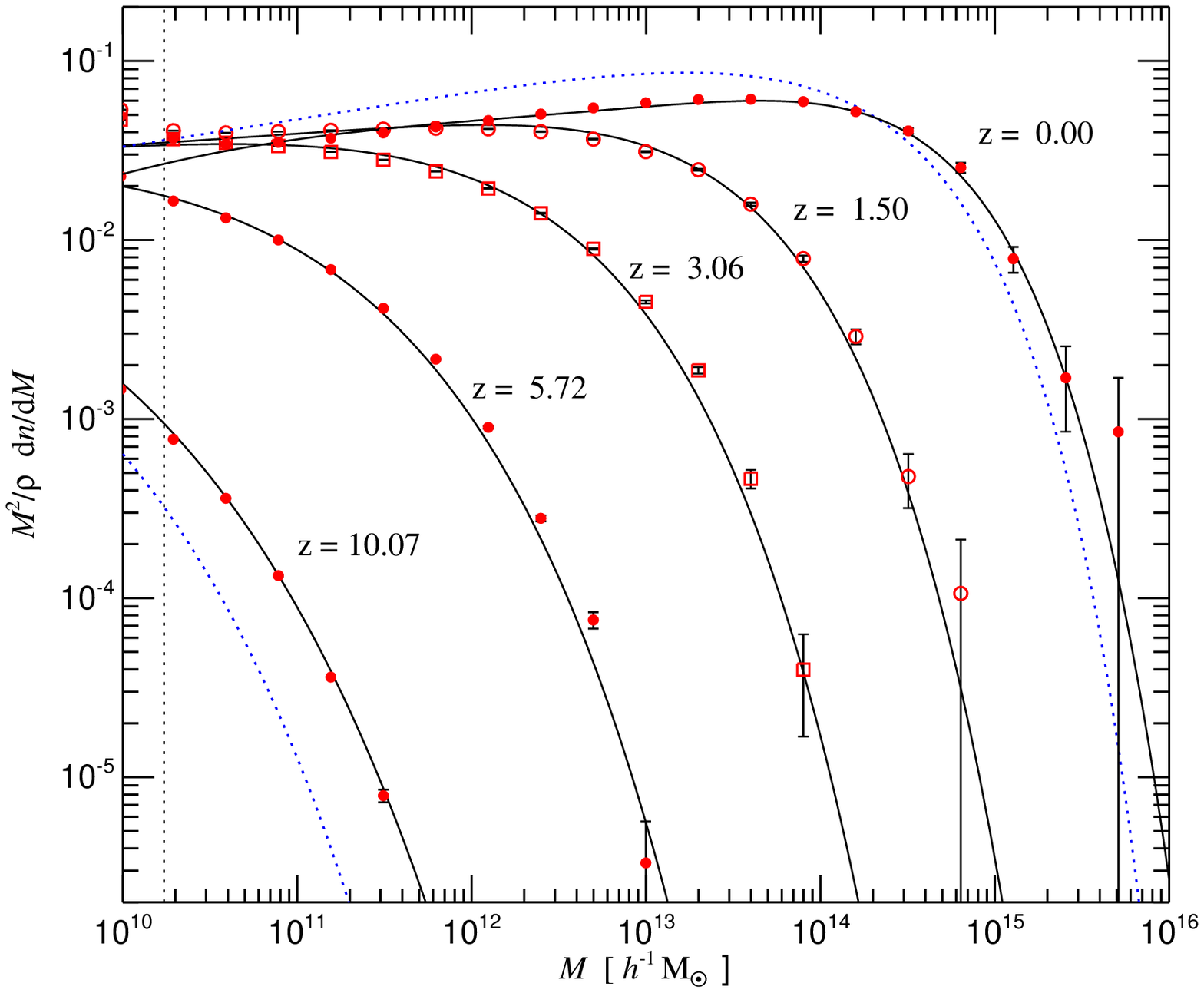,width=6.cm} 
\psfig{file=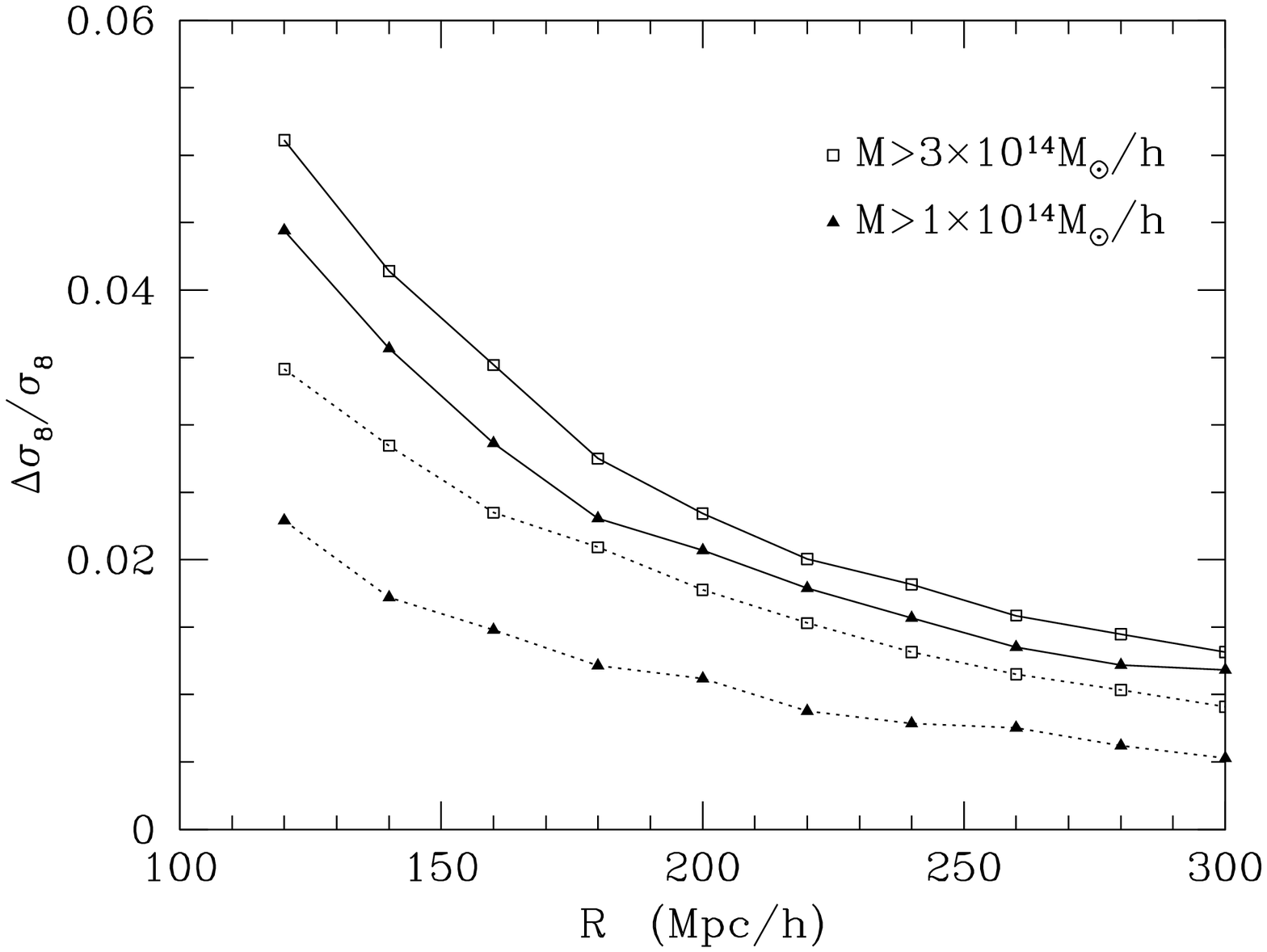,width=6.cm} 
}}
\caption{Left panel: the mass function of DM halos (dots with
  errorbars) identified at different redshifts in the Millenniun Run
  \cite{2005Natur.435..629S}, compared to the predictions of the mass
  function by \cite{JE01.1} and by \cite{PR74.1}. The two model mass
  functions are plotted with solid and dotted curves,
  respectively. Right panel: the relative standard deviation of
  $\sigma_8$ as a function of the sample size
  \cite{2002ApJS..143..241W} from the cumulative mass function above
  two different mass limits. The solid lines indicate the width of the
  distribution when including the clustering of clusters.}
\label{fi:mf_nb} 
\end{figure}

Jenkins et al.  \cite{JE01.1} proposed an alternative expression for
the mass function:
\be
f(\sigma_M,z)=0.315\exp(-|\ln \sigma_M^{-1}+0.61|^{3.8})\,,
\label{eq:jenk}
\ee
which has been obtained as the best fit to the results of a
combination of different simulations, covering a wide dynamical
range. More recently, Springel et al. \cite{2005Natur.435..629S} used
the largest available single N--body simulation to verify in detail
the accuracy of eq.(\ref{eq:jenk}). The result of this comparison,
which is reported in Figure \ref{fi:mf_nb}, demonstrates that this
mass function reproduces remarkably well numerical results over a wide
range of sampled halo masses and redshifts, thereby representing a
substantial improvement with respect to the PS mass function.  The
accuracy of eq.(\ref{eq:jenk}) in reproducing results of numerical
experiments has been also discussed in \cite{2002ApJ...573....7E},
where it is also pointed out the role of different algorithms to
identify clusters and to estimate their mass in simulations, in
\cite{2002ApJS..143..241W}, where the universality of this expression
for a generic cosmology is discussed, and in
\cite{2005astro.ph..6395W}, where the widest dynamical range to date
has been samples by combining a series of N--body simulations.

\begin{figure}
\centerline{
\hbox{\hspace{1.3truecm}\psfig{figure=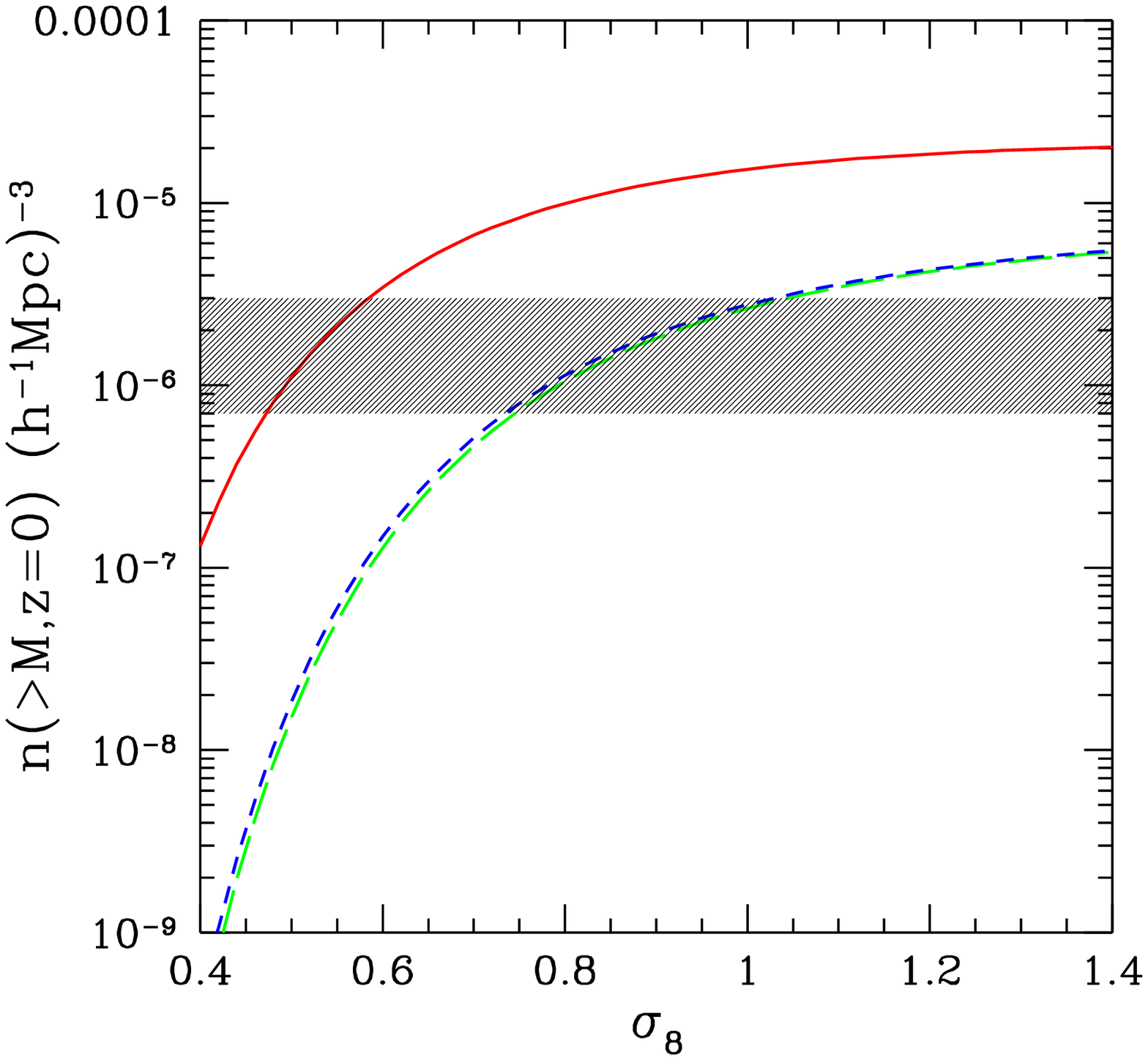,width=6.cm}
\hspace{-0.8truecm}\psfig{figure=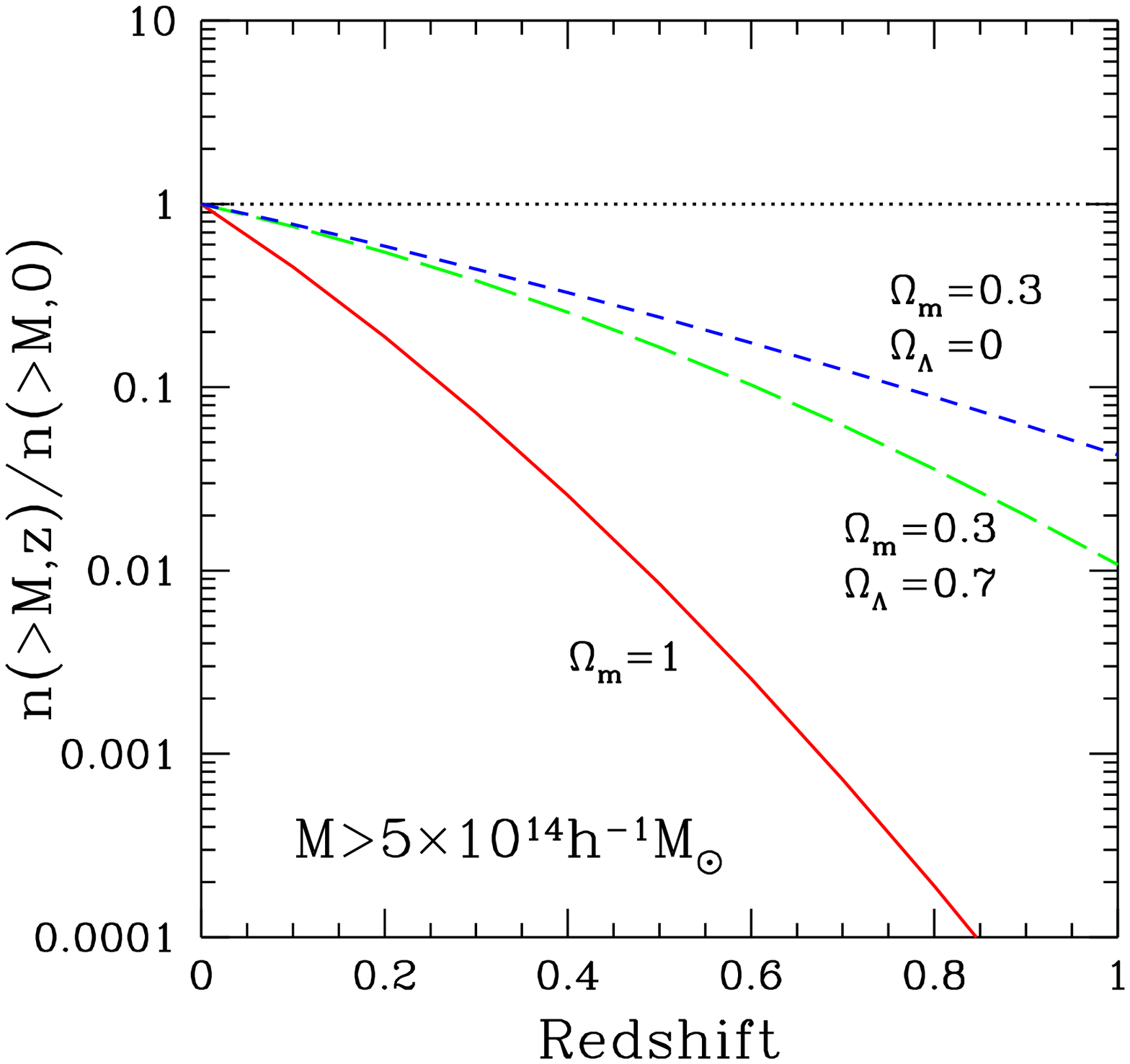,width=6.cm}}
}\null\vspace*{-7mm}
\caption{The sensitivity of the cluster mass function to cosmological
  models \cite{2002ARA&A..40..539R}. Left panel: The cumulative mass
  function at $z=0$ for $M>5\times 10^{14}h^{-1}M_\odot$ for three
  cosmologies, as a function of $\sigma_8$, with shape parameter
  $\Gamma=0.2$; solid line: $\Omega_m=1$; short--dashed line:
  $\Omega_m=0.3$, $\Omega_\Lambda=0.7$; long--dashed line:
  $\Omega_m=0.3$, $\Omega_\Lambda=0$.  The shaded area indicates the
  observational uncertainty in the determination of the local cluster
  space density.  Right panel: Evolution of $n(>\!M,z)$ for the same
  cosmologies and the same mass--limit, with $\sigma_8=0.5$ for the
  $\Omega_m=1$ case and $\sigma_8=0.8$ for the low--density models.}
\label{fi:evol_nm}
\end{figure}

In practical applications, the observational mass function of clusters
is usually determined over about one decade in mass. Therefore, it
probes the power spectrum over a relatively narrow dynamical range,
and does not provide strong constraints on the shape of the power
spectrum. Using only the number density of nearby clusters of a given
mass $M$, one can constrain the amplitude of the density perturbation
at the physical scale $R \propto (M/\Omega_m\rho_{crit})^{1/3}$
which contains this mass. Since such a scale depends both on $M$ and on
$\Omega_m$, the mass function of nearby ($z\mincir 0.1$) clusters is
only able to constrain a relation between $\sigma_8$ and
$\Omega_m$. In the left panel of Figure \ref{fi:evol_nm} we show that,
for a fixed value of the observed cluster mass function, the implied
value of $\sigma_8$ from eq.(\ref{eq:stmf}) increases as the density
parameter decreases.  Determinations of the cluster mass function in
the local Universe using a variety of samples and methods indicate
that $\sigma_8\Omega_m^\alpha\,=\,0.4-0.6$, where $\alpha \simeq
0.4-0.6$, almost independent of the presence of a cosmological
constant term providing spatial flatness. As for the evolution with
redshift, the growth rate of the density perturbations depends primarily on
$\Omega_m$ and, to a lesser extent, on $\Omega_\Lambda$, at least out
to $z\sim 1$, where the evolution of the cluster population is
currently studied. Therefore, following the evolution of the cluster
space density over a large redshift baseline, one can break the
degeneracy between $\sigma_8$ and $\Omega_m$. This is shown in a
pictorial way in Figure~\ref{fi:natfig} and quantified in the right
panel of Figure~\ref{fi:evol_nm}: models with different values of
$\Omega_m$, which are normalized to yield a comparable number density of
nearby clusters, predict cumulative mass functions that progressively
differ by up to orders of magnitude at increasing redshifts.

Although eq.(\ref{eq:jenk}) provides a very accurate and flexible tool
to constrain the parameter space of cosmological models using the mass
function of collapsed halos, nevertheless a further source of
uncertainty may arise from the effect of cosmic variance. Fluctuations
modes, with wavelength exceeding the size of the volumes sampled by
observations, induces appreciable changes in the number counts of
halos of a given mass. This effect has been thoroughly discussed in
\cite{2003ApJ...584..702H,2002ApJS..143..241W}. In the
right panel Figure \ref{fi:mf_nb} (from \cite{2002ApJS..143..241W}) I
report the relative variation of the power spectrum normalization,
$\sigma_8$, induced by cosmic variance, as a function of the sample
size, for halos having two different mass limits. As expected, the
variance decreases with the sample size (fluctuations on larger scales
have a smaller effect), while it increases with the halo mass (the
distribution of rarer objects suffer for a more pronounced
large--scale modulation). This result demonstrates that a precision
calibration of cosmological parameters requires properly accounting
for the effect of cosmic variance.

\section{Building a cluster sample}

\subsection{Identification in the optical / near IR band}
Abell \cite{Ab58} provided the first extensive, statistically complete
sample of galaxy clusters, later extended to the Southern hemisphere
\cite{ACO89}. Based on purely visual inspection, clusters were
identified as enhancements in the galaxy surface density and were
characterized by their {\sl richness} and estimated distance. The
Abell catalog has been for decades the prime source for detailed
studies of individual clusters and for characterizing the large scale
distribution of matter in the nearby Universe. Several variations of
the Abell criteria defining clusters were used in an automated and
objective fashion when digitized optical plates became available
(e.g., \cite{1992MNRAS.258....1L,1997MNRAS.289..263D}). Deep optical
plates were used successfully to search for more distant clusters, out
to $z\simeq 0.9$, with purely visual techniques (e.g.,
\cite{1986ApJ...306...30G,1991MNRAS.249..606C}). These searches for
distant clusters became much more effective with the advent of CCD
imaging. Postman et al. \cite{1996AJ....111..615P} were the first to
carry out a V\&I-band survey over 5 deg$^2$ (the Palomar Distant
Cluster Survey, PDCS). This technique enhances the contrast of galaxy
overdensity at a given position, utilizing prior knowledge of the
luminosity profile typical of galaxy clusters. Dalcanton
\cite{1996ApJ...466...92D} proposed another method of optical
selection of clusters, in which drift scan imaging data from
relatively small telescopes is used to detect clusters as positive
surface brightness fluctuations in the background sky. Gonzalez et
al. \cite{2001ApJS..137..117G} applied a technique based on surface
brightness fluctuations from drift scan imaging data to build a sample
of $\sim\!  1000$ cluster candidates over 130 deg$^2$.

A common feature of all these methods of cluster identification is
that they classify clusters according to definitions of richness,
which generally have a loose relation with the actual cluster
mass. This represents a serious limitation for any cosmological
application, which requires the observable, on which the cluster
selection is based, to be a reliable proxy of the cluster mass.

An improved definition of richness, based on the amplitude of the
galaxy--cluster cross--correlation function, has been applied
\cite{2005ApJS..157....1G} to clusters identified in a large area
survey in $R$ and $z$ bands (the Red Sequence Cluster Survey). This
survey, whose optical and X--ray follow--up, is currently underway,
promises to unveil a fairly large number of clusters out to $z\sim
1.5$.

By increasing the number of observed passbands and using red colors
one can increase the contrast with which clusters are seen in color
space. In this way, one can increase the efficiency of cluster
selection also at high redshift (e.g.,
\cite{2002AJ....123..619S,2005ApJS..157....1G,2005ApJ...634L.129S})
and the accuracy of their estimated redshifts through
spectro--photometric techniques. In this way, Miller et
al. \cite{2005AJ....130..968M} designed a cluster--finding algorithm
which makes full use of information of both position and color space
to detect clusters of galaxies from the SDSS. They were able to
identify about 750 clusters out to $z\mincir 0.2$, and assessed the
degree of completeness by resorting to a comparison with mock SDSS
surveys extracted from large N--body simulations. Once completed, the
search of clusters over the entire SDSS sample will provide about 2500
nearby and medium--distant objects. At the same time the next
generation of wide field ($>\!100$ deg$^2$) deep multicolor surveys in
the optical and especially the near-infrared will powerfully enhance
the search for distant clusters, out to $z\magcir 1$.

\subsection{Identification in the X--ray band}
\label{par:obs}
Already from the first pioneering attempts to map the X--ray sky
(\cite{1972ApJ...178..281G}, see
\cite{2002ARA&A..40..539R} for a historical
review), clusters were associated with extended
sources, whose dominant emission mechanism was recognized to be
thermal bremsstrahlung from optically thin plasma at a temperature of
several keV \cite{1966ApJ...146..955F,1971Natur.231..437C}.  The
all--sky survey conducted by the the HEAO-1 X-ray Observatory was the
first to provide a flux--limited sample of X--ray identified clusters,
for which both the flux number counts and the X--ray luminosity
function have been computed for the first time
\cite{1982ApJ...253..485P}.  However, it is only thanks to the much
improved sensitivity of the {\it Einstein Observatory}
\cite{1979ApJ...230..540G} that X-ray surveys were recognized as an
efficient means of constructing samples of galaxy clusters out to
cosmologically interesting redshifts.

First, the X-ray selection has the advantage of revealing
physically-bound systems, because diffuse emission from a hot ICM is
the direct manifestation of the existence of a potential-well within
which the gas is in dynamical equilibrium with the cool baryonic
matter (galaxies) and the dark matter. Second, the X-ray luminosity is
well correlated with the cluster mass (see Figure~\ref{fi:lm}). Third,
the X-ray emissivity is proportional to the square of the gas density,
hence cluster emission is more concentrated than the optical
bidimensional galaxy distribution. In combination with the relatively
low surface density of X-ray sources, this property makes clusters
high contrast objects in the X-ray sky, and alleviates problems due to
projection effects that affect optical selection. Finally, an inherent
fundamental advantage of X-ray selection is the ability to define
flux-limited samples with well-understood selection functions. This
leads to a simple evaluation of the survey volume and therefore to a
straightforward computation of space densities. Nonetheless, there are
some important caveats described below.  Pioneering work in this field
\cite{1990ApJ...356L..35G,1992ApJ...386..408H} was based on the {\it
Einstein Observatory} Extended Medium Sensitivity Survey (EMSS). The
EMSS survey covered over 700 square degrees and lead to the
construction of a flux-limited sample of 93 clusters out to $z=0.58$,
allowing the cosmological evolution of clusters to be investigated.

The {\it ROSAT} satellite, launched in 1990, allowed a significant
step forward in X-ray surveys of clusters.  The {\it ROSAT} All-Sky
Survey (RASS, \cite{1993Sci...260.1769T}) was the first X-ray
imaging mission to cover the entire sky, thus paving the way to large
contiguous-area surveys of X-ray selected nearby clusters. In the
northern hemisphere, the largest compilations with virtually complete
optical identification include, the Bright Cluster Sample (BCS,
\cite{2000MNRAS.318..333E}), and the Northern {\it ROSAT} All Sky
Survey (NORAS, \cite{2000ApJS..129..435B}). In the southern
hemisphere, the {\it ROSAT}-ESO flux limited X-ray (REFLEX) cluster
survey \cite{2004A&A...425..367B} has completed the identification of
452 clusters, the largest, homogeneous compilation to date. The
Massive Cluster Survey (MACS, \cite{2001ApJ...553..668E}) is aimed
at targeting the most luminous systems at $z>0.3$ which can be
identified in the RASS at the faintest flux levels.  The deepest area
in the RASS, the North Ecliptic Pole (NEP,
\cite{2001ApJ...553L.109H}) which {\it ROSAT} scanned repeatedly
during its All-Sky survey, was used to carry out a complete optical
identification of X-ray sources over a 81 deg$^2$ region. This study
yielded 64 clusters out to redshift $z=0.81$.

In total, surveys covering more than $10^4$ deg$^2$ have yielded over 1000
clusters, out to redshift $z\simeq 0.5$. A large fraction of these are
new discoveries, whereas approximately one third are identified as
clusters in the Abell or Zwicky catalogs.  For the homogeneity of
their selection and the high degree of completeness of their spectroscopic
identifications, these samples are now the basis for a large
number of follow-up investigations and cosmological studies.

Besides the all-sky surveys, the {\it ROSAT-PSPC} archival pointed
observations were intensively used for serendipitous searches of
distant clusters.  These projects, which are now completed, include:
the RIXOS survey \cite{1995Natur.377...39C}, the {\it ROSAT} Deep
Cluster Survey (RDCS,
\cite{1998ApJ...492L..21R,2002ARA&A..40..539R}), the Serendipitous
High-Redshift Archival {\it ROSAT} Cluster survey (SHARC,
\cite{1997ApJ...488L..83B}, the Wide Angle {\it ROSAT} Pointed
X-ray Survey of clusters (WARPS, \cite{2002ApJS..140..265P}), the
160 deg$^2$ large area survey \cite{2003ApJ...594..154M}, the {\it
  ROSAT} Optical X-ray Survey (ROXS, \cite{2001ApJ...552L..93D}).  {\it
  ROSAT}-HRI pointed observations have also been used to search for
distant clusters in the Brera Multi-scale Wavelet catalog (BMW,
\cite{2004A&A...428...21M}).

\begin{figure}
\centerline{
\psfig{figure=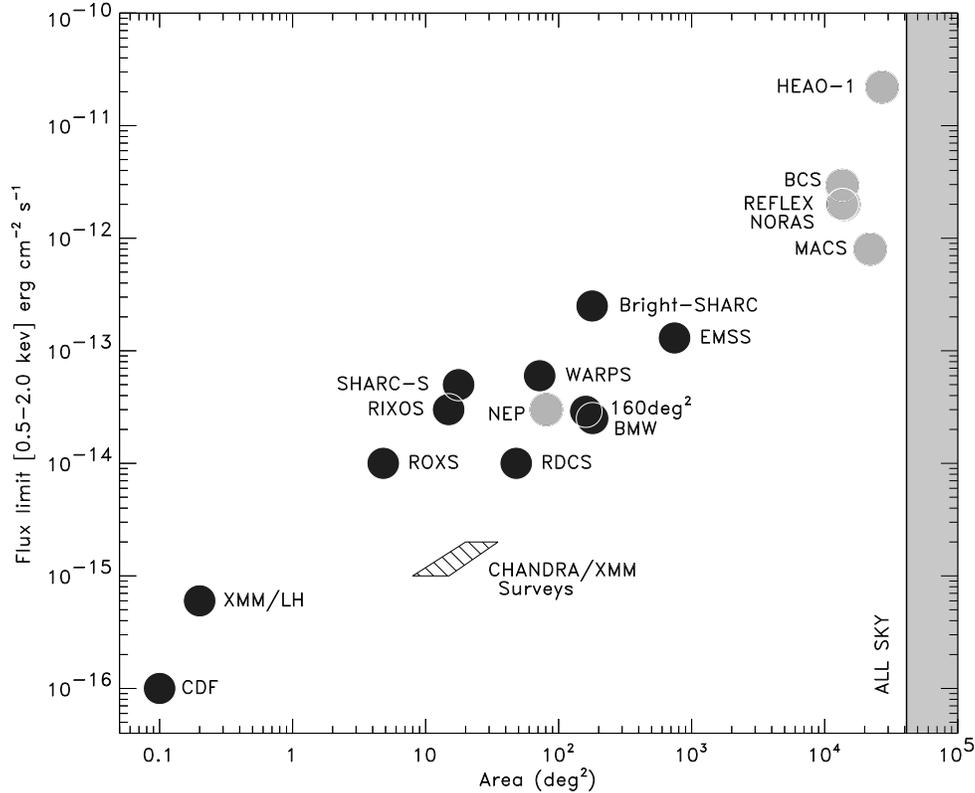,width=13cm}}
\caption{Solid angles and flux limits of X-ray cluster surveys carried
  out over the last two decades. Dark filled circles represent
  serendipitous surveys constructed from a collection of pointed
  observations. Light shaded circles represent surveys covering
  contiguous areas. The hatched region is a predicted locus of current
  serendipitous surveys with {\it Chandra} and {\it Newton-XMM}. From
  \cite{2002ARA&A..40..539R}.}
\label{fi:surveys}
\end{figure}

In Figure~\ref{fi:surveys}, we give an overview of the flux limits and
surveyed areas of all major cluster surveys carried out over the last
two decades.  RASS-based surveys have the advantage of covering
contiguous regions of the sky so that the clustering properties of
clusters (e.g., \cite{2001A&A...368...86S}) can be
investigated. They also have the ability to unveil rare, massive
systems albeit over a limited redshift and X-ray luminosity range.
Serendipitous surveys which are at least a factor of ten deeper but
cover only a few hundreds square degrees, provide complementary
information on lower luminosities, more common systems and are well
suited for studying cluster evolution on a larger redshift baseline.

A number of systematic studies have been carried out to compare the 
nature of clusters identified with the optical and the X--ray technique 
(e.g., \cite{Dona02,Basil04,Plio05}). 
The general conclusion of 
these studies is that optically selected clusters are on average 
underluminous in the X-ray band. This suggests that optical selection 
tends to pick up objects which have not yet reached a high enough 
density to make the ICM lighting up in X--rays.

\begin{figure}
\centerline{
\hbox{\psfig{figure=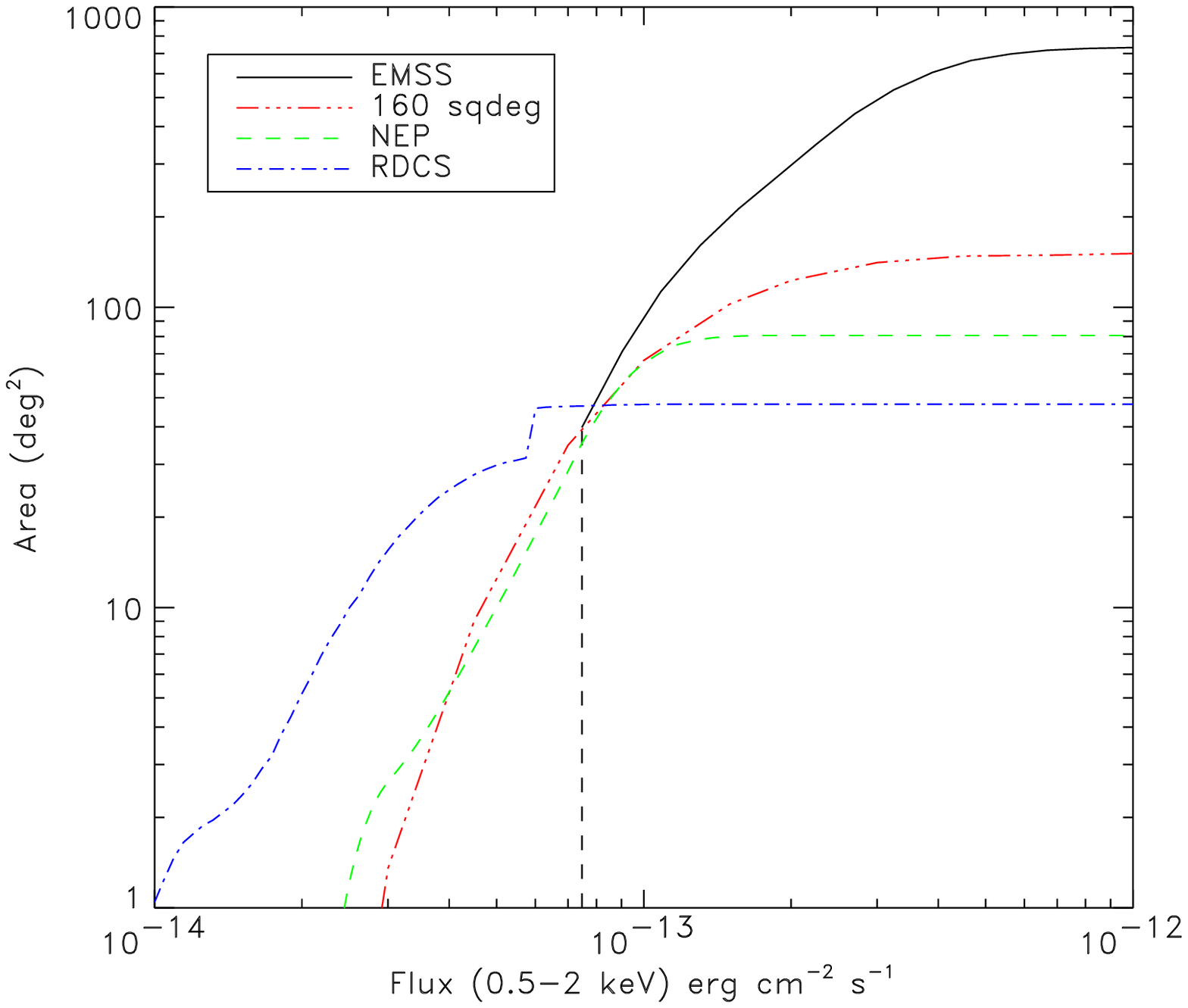,width=6.truecm}
      \psfig{figure=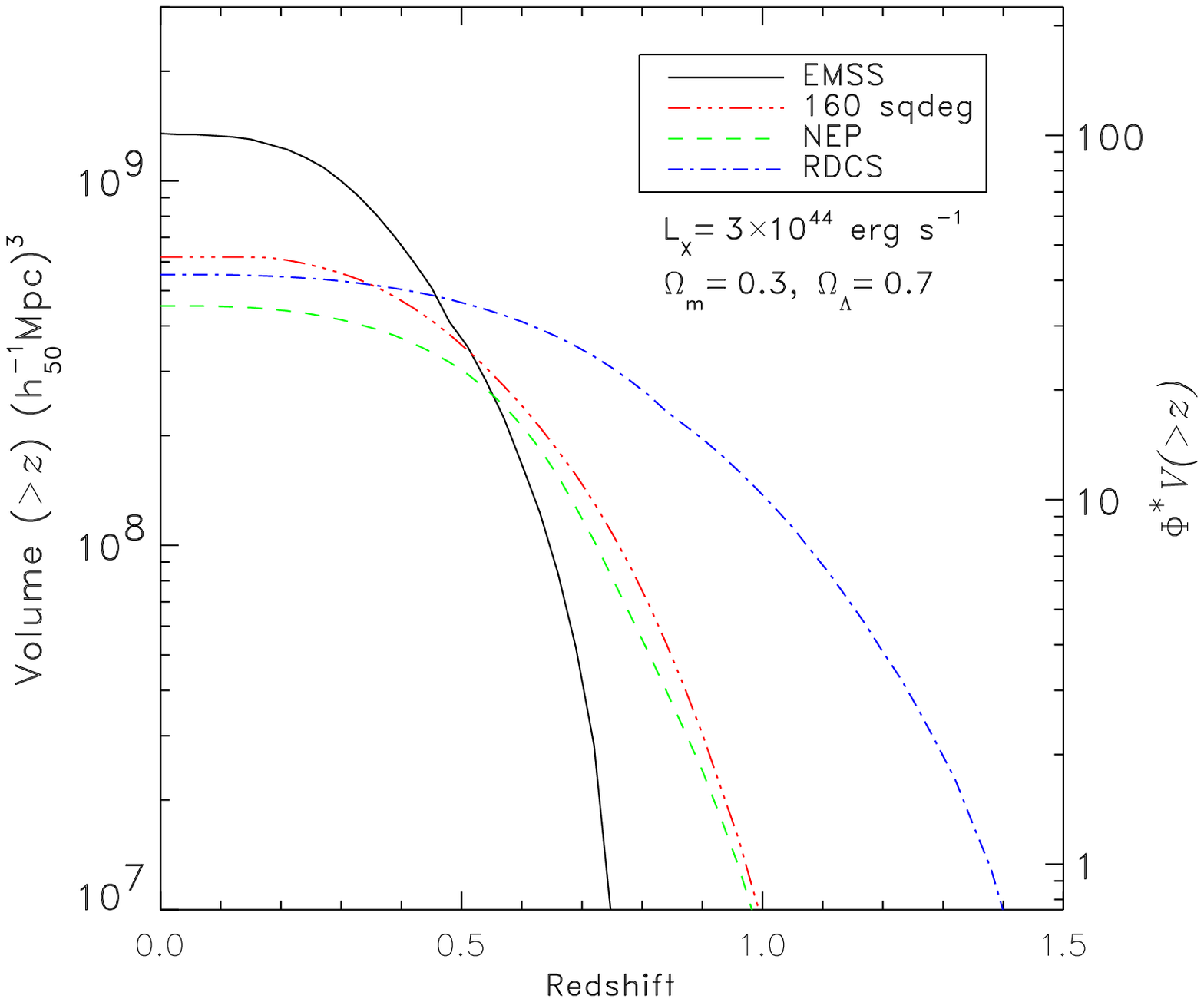,width=6.truecm} 
     }}
   \caption{Left panel: sky coverage as a function of X-ray flux of
     several serendipitous surveys. Right panel: corresponding search
     volumes, $V(>z)$, for a cluster of given $X$-ray luminosity ($L_X
     = 3\times 10^{44}\, [0.5-2\, \rm{keV}]\simeq L_X^* $).  From
     \cite{2002ARA&A..40..539R}.}
\label{fi:vol}
\end{figure}

In order for a survey to be used for cosmological applications, one
needs to know not only how many clusters it contains, but also the
volume within which each of them is found. In other words, one needs
to define the selection function of the survey, which depends on the
survey strategy and on the details of the adopted cluster finding
algorithm (see \cite{2002ARA&A..40..539R}, for a review).  An
essential ingredient for the evaluation of the selection function of
X-ray surveys is the computation of the sky coverage: the effective
area covered by the survey as a function of flux. In general, the
exposure time, as well as the background and the PSF are not uniform
across the field of view of X-ray telescopes, which introduces
vignetting and a degradation of the PSF at increasing off-axis
angles. As a result, the sensitivity to source detection varies
significantly across the survey area so that only bright sources can
be detected over the entire solid angle of the survey, whereas at
faint fluxes the effective area decreases. An example of survey sky
coverage is given in the left panel of Figure~\ref{fi:vol}. By
covering different solid angles at varying fluxes, these surveys probe
different volumes at increasing redshift and therefore different
ranges in X-ray luminosities at varying redshifts.

Once the survey flux--limit and the sky coverage are defined one can
compute the maximum search volume, $V_{max}$, within which a cluster
of a given luminosity is found in that survey:
\be
V_{max}=\int_0^{z_{max}}S[f(L,z)]\left({d_L(z)\over 1+z}\right)^2
{c\,dz\over H(z)}\,.
\label{eq:vmax}
\ee 
Here $S(f)$ is the survey sky coverage, which depends on the flux
$f=L/(4\pi d_L^2)$, $d_L(z)$ is the luminosity distance, and $H(z)$ is
the Hubble constant at $z$. We define $z_{max}$ as the maximum
redshift out to which the flux of an object of luminosity $L$ lies
above the flux limit. The corresponding survey volumes are shown in
the right panel of Figure~\ref{fi:vol}. 

Once again, I emphasize that one of the main advantages of the X--ray
selection lies in the fact that the survey selection function can be
precisely computed, thus allowing reliable comparisons between the
observed and the predicted evolution of the cluster population.

\subsection{Identification through the SZ effect}
\begin{figure}
\centerline{
\hbox{\psfig{figure=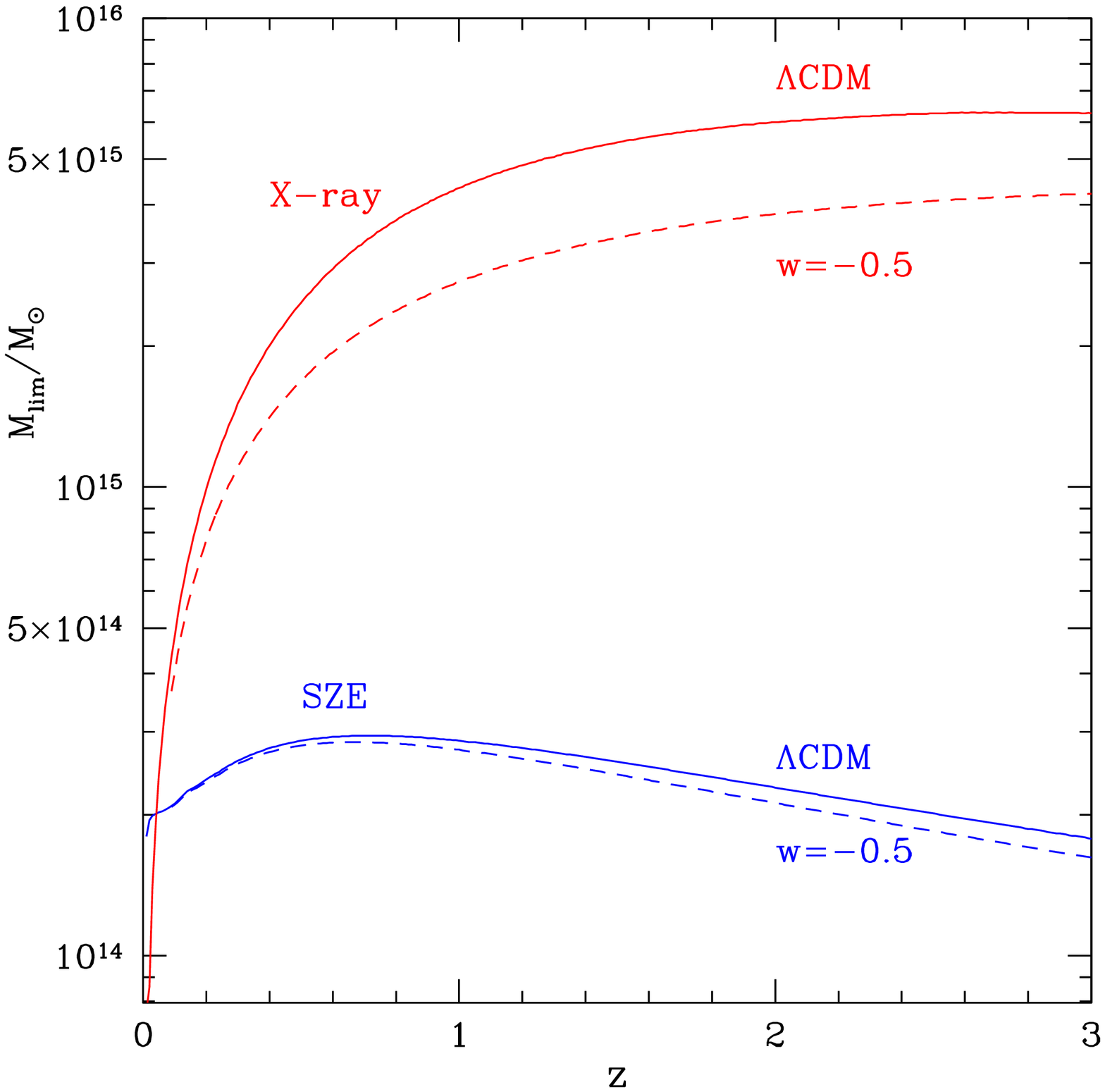,width=6.truecm}
\psfig{figure=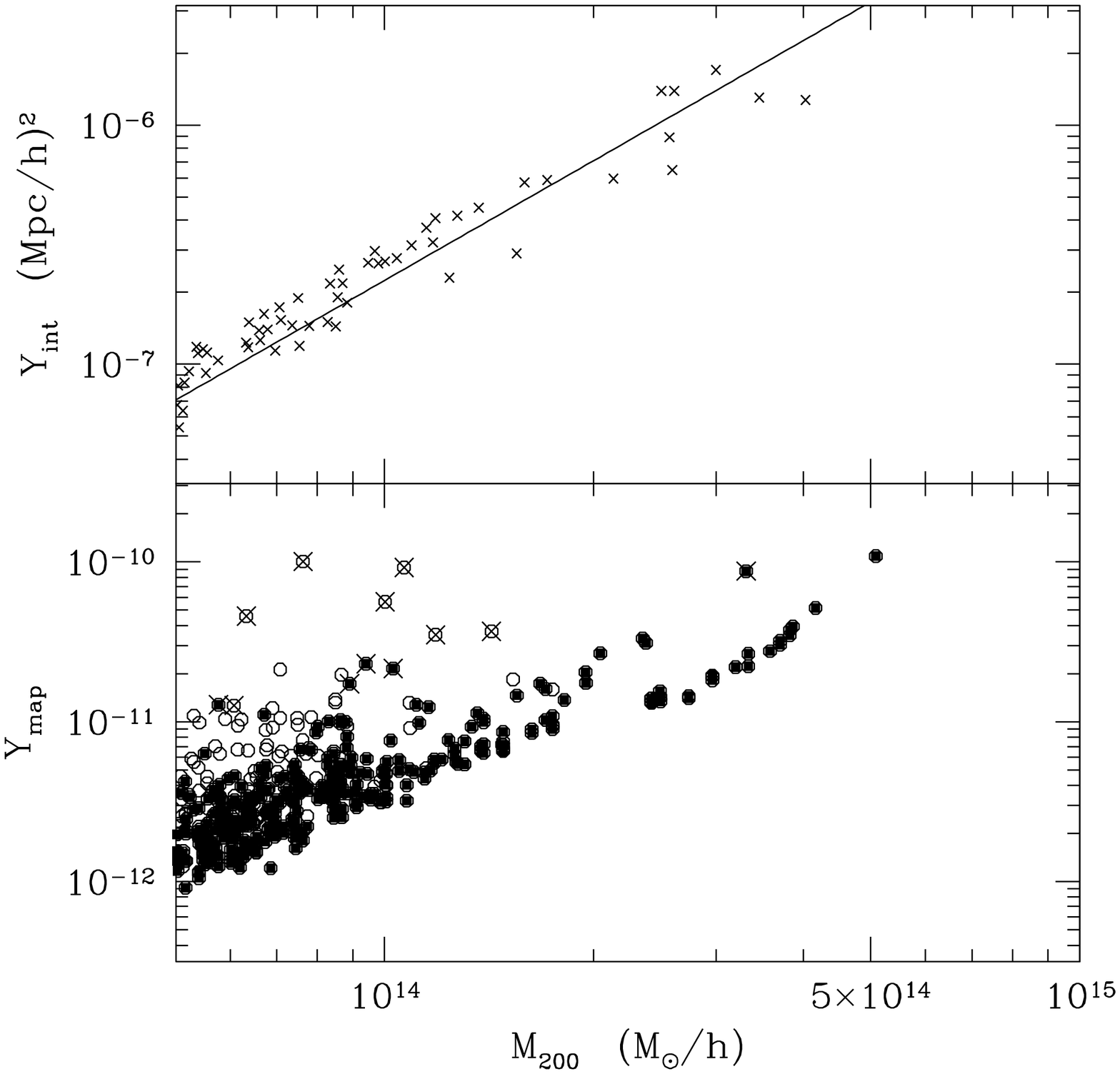,width=6.truecm}
     }}
   \caption{Left panel: limiting cluster virial mass for detection in
     an X--ray and in a SZ survey (from
     \cite{2001ApJ...553..545H}). Each pair of curves show the
     results for two $\Omega_m=0.3$ cosmologies, having $w=-1$ and
     $w=-0.5$ for the DE equation of state. Right panel: the relation
     between the Comptonization parameter and $M_{200}$, from
     \cite{2002ApJ...579...16W}. The upper panel shows the decrement
     contributed from the gas within $0.5R_{200}$. The lower panel
     indicates the signal from noise--free maps projected on the light
     cone.}
\label{fi:szsel}
\end{figure}

The Sunyaev--Zeldovich (SZ) effect \cite{1972CoASP...4..173S} allows
to observe 
galaxy clusters by measuring the distortion of the CMB spectrum
owing to the hot ICM. This method does not depend on redshift and
provides in principle a reliable estimate of cluster masses. For these
reasons, it is now considered as
one of the most powerful means to find distant clusters in
the years to come. For a detailed discussion of the SZ technique for
cluster identification and for the ongoing and future surveys, I refer
to the lectures by Mark Birkinshaw and to the reviews in
\cite{1999PhR...310...97B,2002ARA&A..40..643C}. For the
purpose of the present discussion, I show in the left panel of Figure
\ref{fi:szsel} a comparison between the limiting mass as a function of
redshift, expected for a X--ray and for a SZ cluster survey (from
\cite{2001ApJ...553..545H}). While the standard flux dimming with
the luminosity distance, $f_X\propto d_L^2(z)$, causes the limiting
mass to quickly increase with distance for the X--ray selection, this
limiting mass has a much less sensitive dependence on redshift for the
SZ selection. This is the reason why SZ surveys are generally
considered as essentially providing mass--limited cluster samples.

It has been recently pointed out \cite{2005ApJ...623L..63M} that the
integrated SZ flux--decrement has a very tight correlation with the
total cluster mass (see also \cite{2005MNRAS.356.1477D}). This fact,
joined with the redshift--independence of the SZ selection, makes the
SZ identification a promising route toward precision cosmology with
galaxy clusters.

A potential problem with the SZ identification of clusters resides in
the possible contamination of the signal from foreground/background
structures. Diffuse gas, residing in large--scale filaments, are
likely to provide a negligible contamination, as a consequence of the
comparatively low density and temperature which characterize such
structures. However, small halos, which are expected to be present in
large number, contain gas at the virial overdensity. Since they are
not resolved in current SZ observations, their integrated contribution
may provide a significant contamination. Using cosmological
hydro-dynamical simulations, White et al. \cite{2002ApJ...579...16W}
have created SZ sky maps with the aim of correlating the SZ signal
seen in projection with the actual mass of clusters.The result of this
test is shown in the right panel of Fig. \ref{fi:szsel}. The upper
panel shows the relation between the integrated SZ signal contributed
only from the gas within $0.5R_{200}$ and $M_{200}$, while the lower
panel is when using the actual Compton--$y$ parameter measured from
the projected maps. Quite apparently, the scatter in the relation is
significantly increased in projection. Part of the scatter is due to
the different redshifts at which clusters seen in projection are
placed. This contribution to the scatter can be removed once redshifts
of clusters are known from follow--up optical observations. However, a
significant contribution to the overall scatter is contributed by
cluster asphericity and by contamination from fore/background
structures. This highlights the relevance of keeping this scatter
under control for a full exploitation of the SZ signal as a tracer of
the cluster mass.

\section{Methods to estimate cluster masses}

\subsection{The hydrostatic equilibrium}
\label{s:he}
The condition of hydrostatic equilibrium determines the balance
between the pressure force and the gravitational force: $\nabla
P_{\rm gas}=-\rho_{\rm gas}\nabla \phi$, where $P_{\rm gas}$ and
$\rho_{\rm gas}$ are the gas pressure and density, respectively, while
$\phi$ is the underlying gravitational potential. Under the assumption
of a spherically symmetric gas distribution, the above equations read:
\be
{dP_{\rm gas}\over dr}=-\rho_{\rm gas}{d\phi\over dr}=-\rho_{\rm
  gas}{GM(<r)\over r^2}\,,
\ee
where $r$ is the radial coordinate (clustercentric distance) and
$M(<r)$ is the total mass contained within $r$. Using the equation of
state of ideal gas to relate pressure to gas density and temperature,
the mass is then given by
\be
M(<r)=-{r\over G}{k_BT\over \mu m_p}\left({d\ln \rho_{\rm gas} \over
    d\ln r}+{d\ln T\over d\ln r}\right)\,,
\label{eq:mhe}
\ee
where $\mu$ is the mean molecular weight of the gas ($\mu\simeq 0.59$
for primordial composition) and $m_p$ is the proton mass. An often
used mass estimator is based on assuming the $\beta$--model for the
gas density profile,
\be
\rho_{\rm gas}(r)={\rho_0\over \left[ 1+(r/r_c)^2\right]^{3\beta /2}}
\label{eq:betam}
\ee
\cite{1976A&A....49..137C}. In the above equation, $r_c$ is the core
radius, while $\beta$ is the ratio between the kinetic energy of any
tracer of the gravitational potential (e.g. galaxies) and the thermal
energy of the gas, $\beta=\mu m_p\sigma_v^2/(k_BT)$ ($\sigma_v$:
one--dimensional velocity dispersion). By further assuming a
polytropic equation of state, $\rho_{\rm gas}\propto P_{\rm
  gas}^\gamma$ ($\gamma$: polytropic index), eq.(\ref{eq:mhe}) becomes
\be
M(<r)\simeq 1.11 \times 10^{14}\beta \gamma {T(r)\over {\rm
    keV}}{r\over \hm}{(r/r_c)^2\over 1+(r/r_c)^2}\msun\,,
\label{eq:begam}
\ee
where $T(r)$ is the temperature at the radius $r$.  In its original
derivation, the $\beta$--model was aimed at representing the
distribution of isothermal gas sitting in hydrostatic equilibrium
within a King--like potential. The corresponding mass estimator is
recovered from eq.(\ref{eq:begam}) by setting $\gamma=1$ and replacing
$T(r)$ with the global ICM temperature, $T_0$. In the absence of
accurately resolved temperature profiles from X--ray observations,
eq.(\ref{eq:begam}) has been used to estimate cluster masses both in
its isothermal (e.g., \cite{RE02.1}) and in its polytropic form
(e.g., \cite{2000ApJ...532..694N,2001A&A...368..749F,2002A&A...391..841E}).

Thanks to the much improved sensitivity of the Chandra and XMM--Newton
X--ray observatories, temperature profiles are now resolved with high
enough accuracy to allow the application of more general methods of
mass estimation, not necessarily bound to the assumptions of
$\beta$--model and of an overall polytropic form for the equation of
state (e.g., \cite{2001MNRAS.328L..37A,2002A&A...391..841E,2005A&A...441..893A,2005astro.ph..7092V}).

\begin{figure}
\centerline{\hbox{
%\vspace{-2.cm}
\psfig{figure=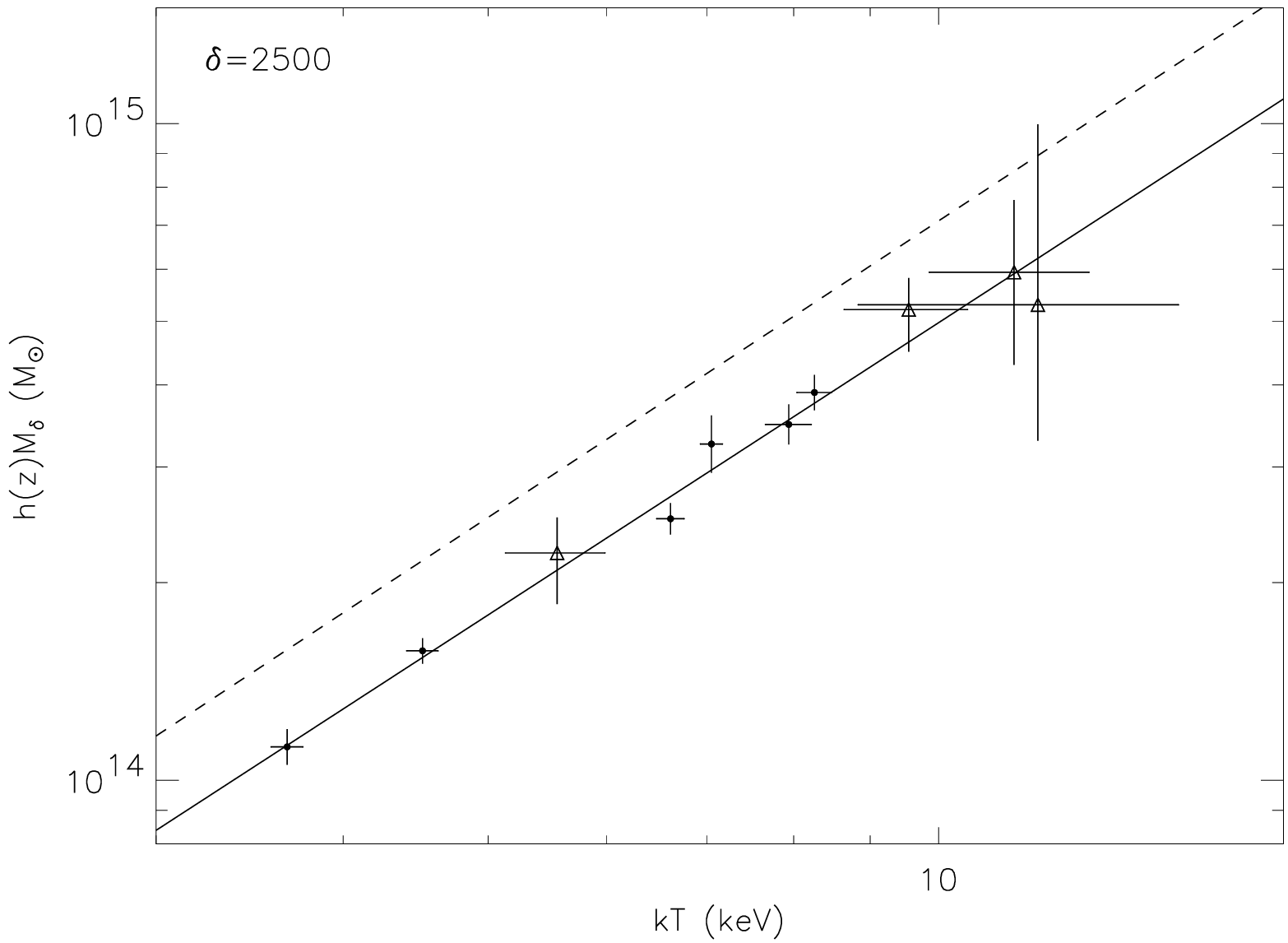,width=6.5cm}
\psfig{figure=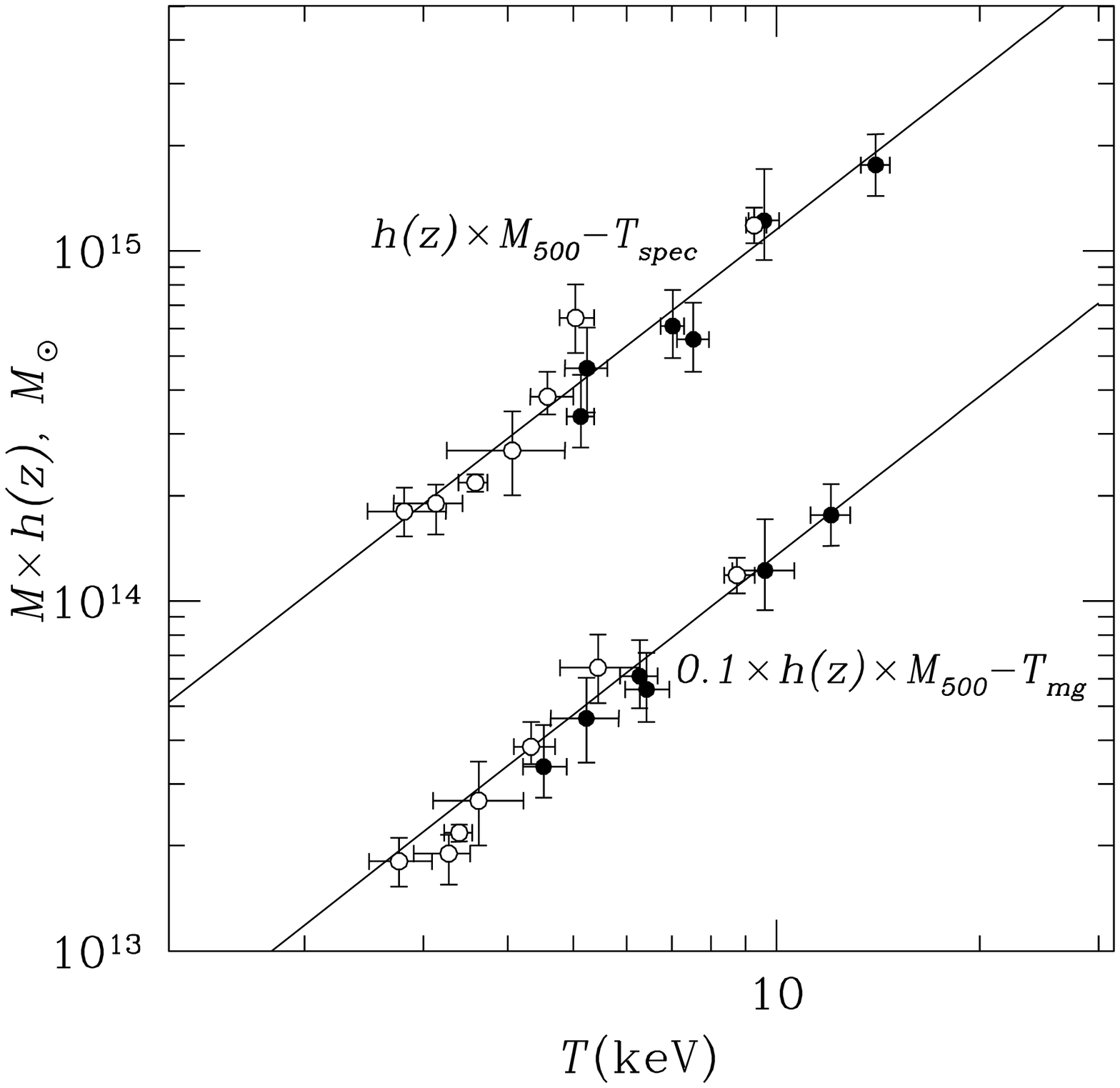,width=5.5cm} 
     }}
   \caption{The mass-temperature relation for nearby clusters (from
     \cite{2005A&A...441..893A}) and for distant clusters (from
     \cite{2005ApJ...633..781K}), based on a combination of Chandra
     and XMM--Newton data.}
\label{fi:mt}
\end{figure}

An alternative way of recasting the isothermal version of
eq.(\ref{eq:begam}) between temperature and mass is based on
expressing the mass according to the virial theorem as $M_{\rm vir} =
\sigma_v^2 R_{\rm vir}/G$, so that
\be
k_BT={1.38\over \beta}\left( {M_{\rm vir} \over
    10^{15}\msun}\right)^{3/2}\left[ \Omega_m \Delta_{\rm
    vir}(z)\right]^{1/3} (1+z)\,{\rm keV}\,.
\label{eq:mteke} 
\ee
This expression, originally introduced in \cite{1996MNRAS.282..263E},
has been sometimes used to express the $M$--$T$ relation as obtained
from hydrodynamical simulations of galaxy clusters (e.g., 
\cite{1998ApJ...495...80B,2002MNRAS.336..409B}).

It is clear that the two crucial assumptions underlying any mass
measurements based on the ICM temperature concerns the existence of
hydrostatic equilibrium and of spherical symmetry. While effects of
non--spherical geometry can be averaged out by performing the analysis
over a large enough number of clusters, the former can lead to
systematic biases in the mass estimates (e.g.,
\cite{2005ApJ...618L...1R}) and references therein). So far, ICM
temperature measurements have been based on fits of the observed
X--ray spectra of clusters to plasma models, which are dominated at
high temperatures by thermal bremsstrahlung. However, local deviations
from isothermality, e.g. due to the presence of merging cold gas
clumps, can bias the spectroscopic temperature with respect to the
actual electron temperature (e.g.,
\cite{2001ApJ...546..100M,2004MNRAS.354...10M,2005astro.ph..4098V}). This
bias directly translates into a comparable bias in the mass estimate
through hydrostatic equilibrium (see Section \ref{s:fut}, below).

\subsection{The dynamics of member galaxies}
From a historical point of view, the dynamics traced by member
galaxies, has been the first method applied to measure masses of
galaxy clusters \cite{1936ApJ....83...23S,1937ApJ....86..217Z}. Under
the assumption of virial equilibrium, the mass of the cluster can be
estimated by knowing position and redshift for a high enough number of
member galaxies:
\be 
M={\pi\over 2}{3\sigma_v^2 R_V\over G}
\label{eq:mvir}
\ee
(e.g., \cite{1960ApJ...132..286L}), where the first factor accounts for the
geometry of projection, $\sigma_r$ is the line-of-sight velocity
dispersion and $R_V$ is the virialization radius, which depends on the
positions of the galaxies with measured redshifts and recognized as
true cluster members:
\be
R_V=N^2\left( \sum_{i>j}r_{ij}^{-1}\right)^{-1}\,,
\ee
where $N$ is the total number of galaxies, and $r_{ij}$ the projected
separation between the $i$-th and $j$-th galaxies. This method has
been extensively applied to measure masses for statistical samples of
both nearby (e.g.,
\cite{1993ApJ...411L..13B,1998ApJ...505...74G,2003ApJ...585..205B,2003AJ....126.2152R,2005A&A...433..431P})
and distant (e.g., \cite{1997ApJ...476L...7C,2001ApJ...548...79G})
clusters.

Besides the assumption of virial equilibrium, which may be fulfilled
to different degrees by different populations of galaxies (e.g., late
vs. early type), a crucial aspect in the application of the dynamical
mass estimator concerns the rejection of interlopers, i.e. of
back/foreground galaxies which lie along the line-of-sight of the
cluster without belonging to it. A spurious inclusion of non--member
galaxies in the analysis leads in general to an overestimate of the
velocity dispersion and, therefore, of the resulting mass. A number of
algorithms have been developed for interlopers rejection, whose
reliability must be judged on a case-by-case basis (e.g., 
\cite{1993ApJ...404...38G,1997MNRAS.287..817V}). A further potential
problem of this analysis concerns the possibility of realizing a
uniform sampling of the cluster potential using galaxies with measured
redshifts. For instance, the technical difficulty of packing slits or
fibers in optical spectroscopic observations may lead to an
undersampling of the cluster central regions. In turn, this leads to
an overestimate of $R_V$ and, again, of the collapsed mass.

Tests of the accuracy of mess estimates based on the dynamical virial
method have been performed by using hydrodynamical simulations of
galaxy clusters, in which galaxies are identified from gas cooling and
star formation (\cite{1996ApJ...472..460F,Biviano06}).
For instance, \cite{Biviano06} have shown that galaxies identified in the
simulations are fair tracers of the underlying dynamics, with
no systematic bias in the estimate of cluster masses, although a
rather large scatter between true and recovered masses is induced
mostly by projection effects.

\begin{figure}
\centerline{\hbox{
\psfig{figure=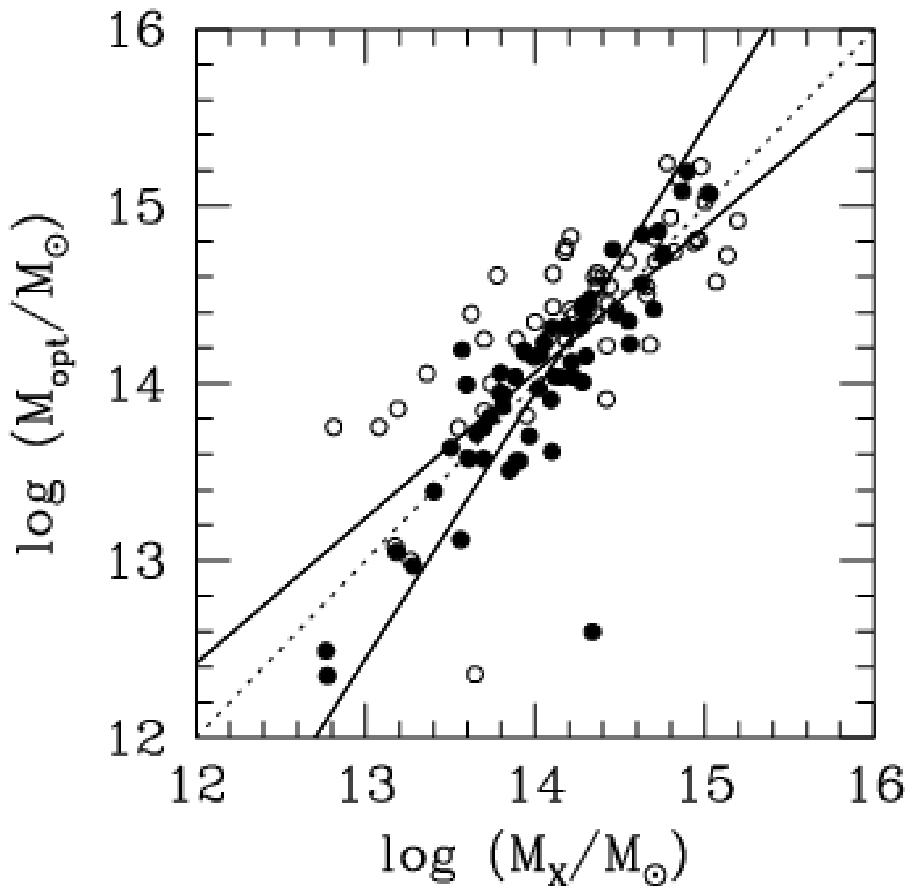,width=6.5cm}
\psfig{figure=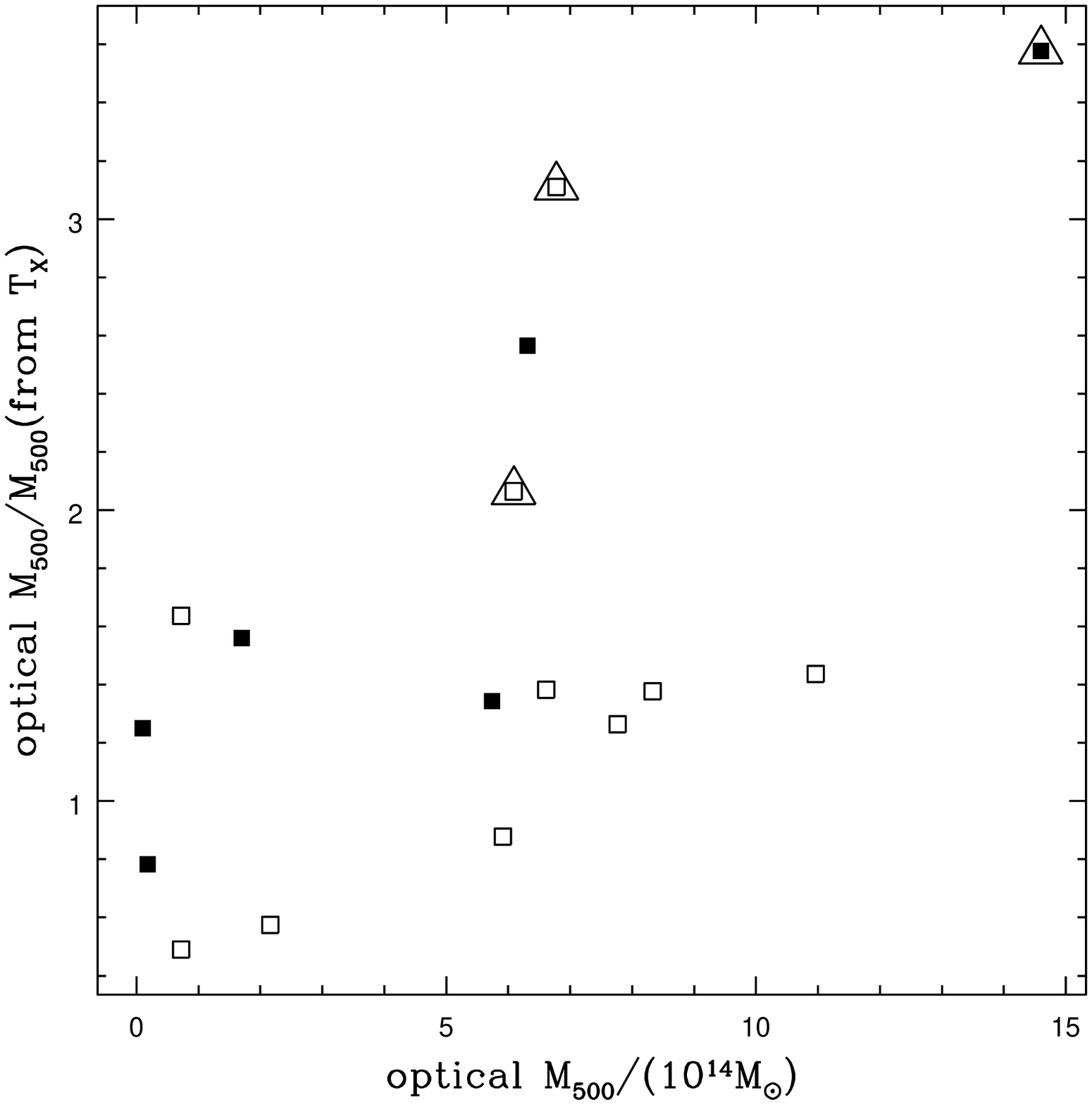,width=5.cm,bbllx=50bp,bblly=50bp,bburx=600bp,bbury=600bp}
     }}
   \caption{The relation between dynamical optical masses and masses
     derived from the X--ray temperature by assuming hydrostatic
     equilibrium (from \cite{1998ApJ...505...74G}, left panel, and
     from \cite{2005A&A...433..431P}, right panel, based on SDSS
     spectroscopic data).}
\label{fi:mxmo}
\end{figure}

Quite reassuringly, despite all the assumptions and possible
systematics affecting both dynamical optical and X--ray mass
estimates, these two methods provide in general fairly consistent
results for both nearby (e.g., 
\cite{1998ApJ...505...74G,2005A&A...433..431P}) and distant (e.g., 
\cite{1999ApJ...517..587L}) clusters. Two examples of such comparisons
are shown in Figure \ref{fi:mxmo}. In the left panel, we report the
comparison between X--ray and optical dynamical masses
\cite{1998ApJ...505...74G}. This plot shows a reasonable agreement among
the two mass estimates, although with some scatter. The right panel
reports the comparison presented in \cite{2005A&A...433..431P}. In this plot,
the triangles indicates the cluster with clear evidences of complex
dynamics. Quite interestingly, the agreement between the two mass
estimates is acceptable, with a few outliers which are generally
identified with non--relaxed clusters.

\subsection{The self--similar scaling}
The simplest model to explain the physics of the ICM is based on the
assumption that gravity only determines the thermodynamical properties
of the hot diffuse gas \cite{1986MNRAS.222..323K}. Since gravity does
not have a preferred scale, we expect clusters of different sizes to
be the scaled version of each other as long as gravity only determines
the ICM evolution and there are no preferred scales in the underlying
cosmological model. This is the reason why the ICM model based on the
effect of gravity only is said to be self-similar.

If we define $M_{\Delta_c}$ as the mass contained within the radius
$R_{\Delta_c}$, encompassing a mean density $\Delta_c$ times the
critical density, then $M_{\Delta_c} \propto \rho_c(z) \Delta_c
R_{\Delta_c}^3 $. Here $\rho_c(z)$ is the critical density of the
universe which scales with redshift as $\rho_c(z)=\rho_{c,0} E^2(z)$,
where $E(z)$ is given by eq.(\ref{eq:ez}). On the other hand, the
cluster size $R$ scales with $z$ and $M_{\Delta_c}$ as $R\propto
M^{1/3} E^{-2/3}(z)$. Therefore, assuming hydrostatic equilibrium, the
cluster mass scales with the temperature $T$ as
\begin{equation}
M_{\Delta_c}\,\propto \,T^{3/2}E^{-1}(z)\,.
\label{eq:mt_ss}
\end{equation}
If $\rho_{\rm gas}$ is the gas density, the corresponding 
X--ray luminosity for pure thermal bremsstrahlung emission is
\begin{equation}
L_X\,=\, \int_V \left({\rho_{\rm gas}\over \mu m_p}\right)^2
\Lambda(T)\,dV\,, 
\label{eq:lx}
\end{equation}
where $\Lambda(T)\propto T^{1/2}$. Further assuming that the gas
distribution traces the dark matter distribution, $\rho_{\rm
gas}(r)\propto \rho_{DM}(r)$, then
\begin{equation}
L_X\,\propto\,M_{\Delta_c}\rho_c T^{1/2}\,\propto \, T^2 E(z)\,.
\label{eq:lt_ss}
\end{equation}

As for the CMB intensity decrement due to the thermal SZ effect we
have 
\begin{equation}
\Delta S\, \propto\, \int y(\theta) {\rm d}\Omega\, \propto
\,d_A^{-2} \int T n_e {\rm d}^3r \propto d_A^{-2} T^{5/2} E^{-1}(z)\,,
\label{eq:ST_ss}
\end{equation}
where $y$ is the Comptonization parameter, $d_A$ is the angular size
distance and $n_e$ is the electron number density. We can also write
$\Delta S$ in a different way to get the explicit dependence on $y_0$:
\begin{equation}
\Delta S\, \propto \,y_0 d_A^{-2}
\int {\rm d}\Omega \propto y_0 d_A^{-2} M^{2/3} E^{-4/3}(z)\, \propto
\,y_0 d_A^{-2} T E^{-2}(z)\,. 
\end{equation}
In this way, we obtain the following scalings for the central value of
the Comptonization parameter:
\begin{equation}
y_0 \,\propto \,T^{3/2}E(z) \,\propto \, L_X^{3/4} E^{1/4}(z)\,.
\label{eq:yt_ss}
\end{equation}

Eqs.(\ref{eq:mt_ss}), (\ref{eq:lt_ss}) and (\ref{eq:yt_ss}) are unique
predictions for the scaling relations among ICM physical quantities
and, in principle, they provide a way to relate the cluster masses to
observables at different redshifts. As we shall discuss in the
following, deviations with respect to these relations witness the
presence of more complex physical processes, beyond gravitational
dynamics only, which affect the thermodynamical properties of the
diffuse baryons and, therefore, the relation between observables and
cluster masses.

\subsection{Phenomenological scaling relations}

\subsubsection{Using the X--ray luminosity}
\begin{figure}
\centerline{\hbox{
\psfig{figure=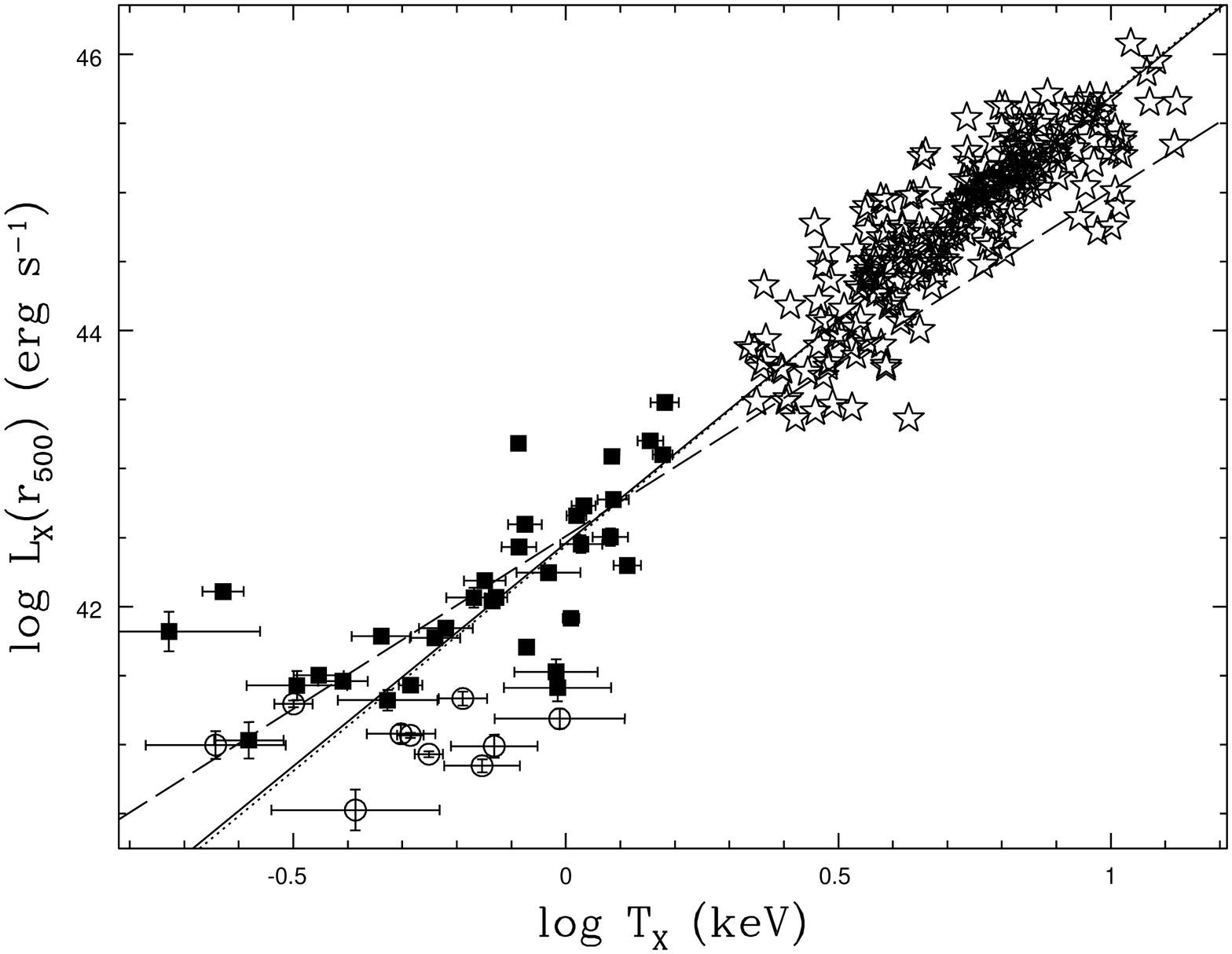,angle=270,width=6.cm}
\psfig{figure=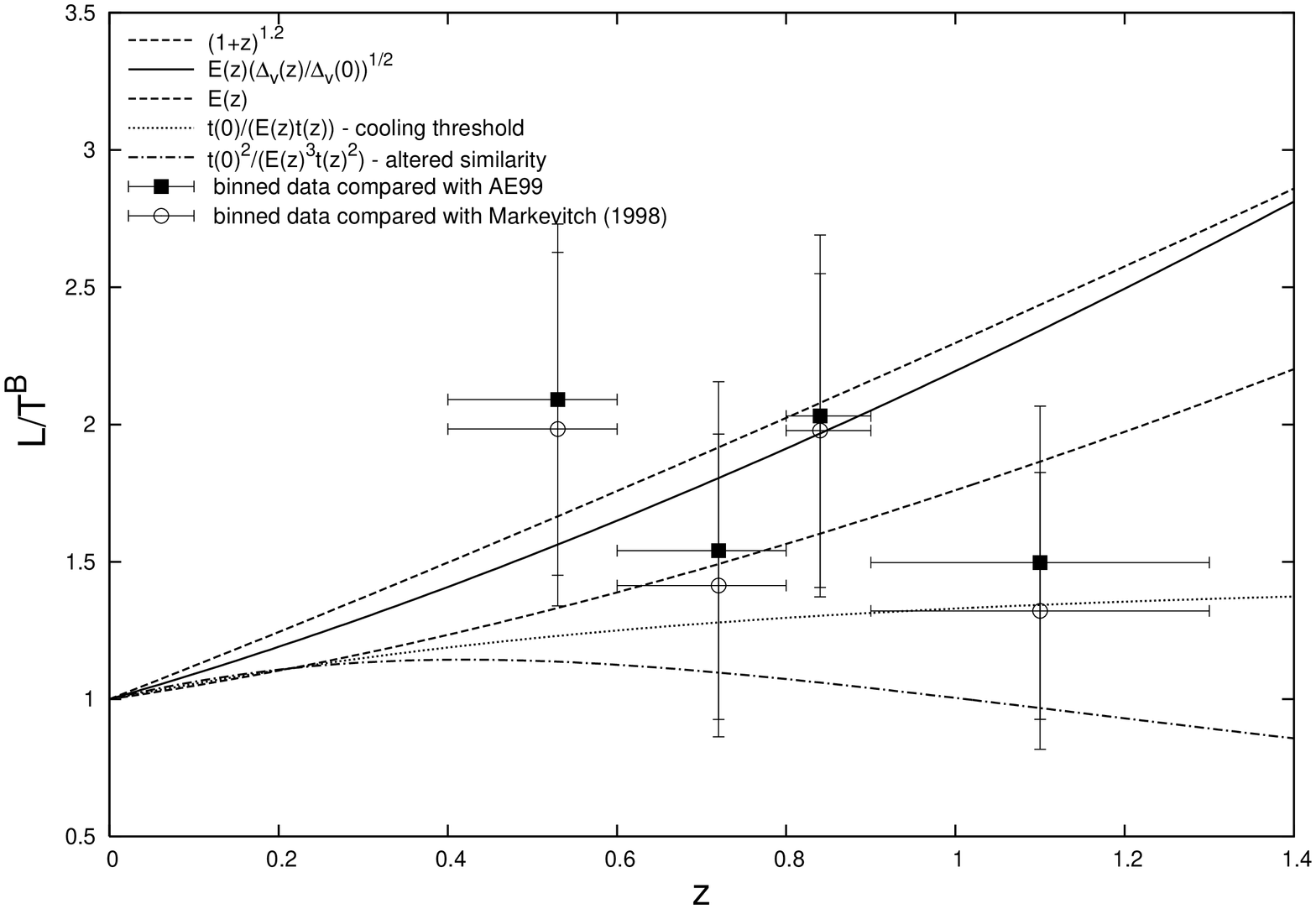,angle=270,width=6.cm} 
     }}
   \caption{Left panel: the $L_X$--$T$ relation for nearby clusters
     and groups (from \cite{2004MNRAS.350.1511O}). The star symbols
     are the for the sample of clusters, with temperature measured
     from ASCA, while the filled squares and open circles are for a
     sample of groups, also with ASCA temperatures. Right panel: the
     evolution of the $L_X$--$T$ relation, normalized to the local
     relation (from \cite{2006MNRAS.365..509M}), using Chandra
     temperatures of clusters at $z>0.4$.}
\label{fi:lt}
\end{figure}

The relation between X--ray luminosity and temperature of nearby
clusters is considered as one of the most robust observational facts
against the self--similar model of the ICM. A number of observational
determinations now exist, pointing toward a relation $L_X\propto
T^\alpha$, with $\alpha\simeq 2.5$--3 (e.g.,
\cite{2000ApJ...538...65X}), possibly flattening towards the
self--similar scaling only for the very hot systems with $T\magcir 10$
keV \cite{Allen98}. While in general the scatter around the
best--fitting relation is non negligible, it has been shown to be
significantly reduced after excising the contribution to the
luminosity from the cluster cooling regions \cite{1998ApJ...504...27M}
or by removing from the sample clusters with evidence of cooling flows
\cite{1999MNRAS.305..631A}. As for the behaviour of this relation at
the scale of groups, $T\mincir 1$ keV, the emerging picture now is
that it lies on the extension of the $L_X$--$T$ relation of clusters,
with no evidence for a steepening \cite{1998ApJ...496...73M}, although
with a significant increase of the scatter \cite{2004MNRAS.350.1511O},
possibly caused by a larger diversity of the groups population when
compared to the cluster population.  This result is reported in the left
panel of Figure \ref{fi:lt} (from \cite{2004MNRAS.350.1511O}), which
shows the $L_X$--$T$ relation for a set of clusters with measured ASCA
temperatures and for a set of groups.

As for the evolution of the $L_X$--$T$ relation, a number of analyses
have been performed, using Chandra
\cite{2002AJ....124...33H,2002ApJ...578L.107V,2004A&A...417...13E,2006MNRAS.365..509M}
and XMM--Newton \cite{2005ApJ...633..781K,2004A&A...420..853L}
data. Although some differences exist between the results obtained
from different authors, such differences are most likely due to the
convention adopted for the radii within which luminosity and
temperature are estimated. In general, the emerging picture is that
clusters at high redshift are relatively brighter, at fixed
temperature. The resulting evolution for a cosmology with
$\Omega_m=0.3$ and $\Omega_\Lambda=0.7$ is consistent with the
predictions of the self--similar scaling, although the slope of the
high--$z$ $L_X$--$T$ relation is steeper than predicted by
self--similar scaling, in keeping with results for nearby
clusters. The left panel of Fig. \ref{fi:lt} shows the evolution of
the $L_X$--$T$ relation from \cite{2006MNRAS.365..509M}, where Chandra
and XMM--Newton observations of 11 clusters with redshift $0.6<z<1.0$
were analyzed. The vertical axis reports the quantity $L_X/T^B$, where
$B$ is the slope of the local relation. Quite apparently, distant
clusters are systematically brighter relatively to the local
ones. However, the uncertainties are still large enough not to allow
the determination of a precise redshift dependence of the $L_X$--$T$
normalization.

\begin{figure}
\centerline{
\psfig{figure=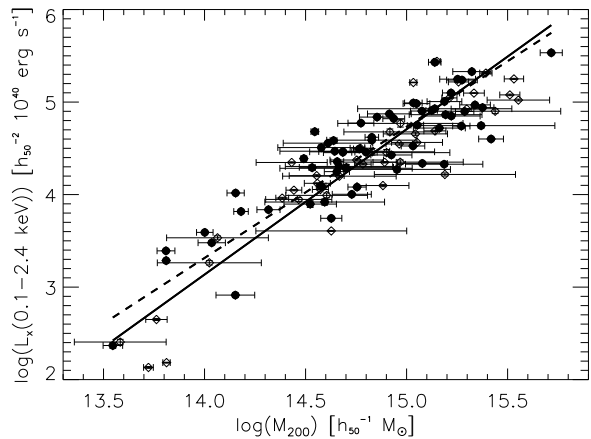,width=9.cm}}
\caption{The $L_X$--$M$ relation for nearby clusters (from
  \cite{RE02.1}). $X$--ray luminosities are from the RASS,
  while masses are estimated using ASCA temperatures and assuming
  hydrostatic equilibrium for isothermal gas.}
\label{fi:lm}
\end{figure}

As for the relation between X--ray luminosity and mass, its first
calibration has been presented in \cite{RE02.1}, for a
sample of bright clusters extracted from the ROSAT All Sky Survey
(RASS). In their analysis, these authors derived masses by using
temperatures derived from ASCA observations and applying
the equation of hydrostatic equilibrium, eq.(\ref{eq:mhe}), for an isothermal
$\beta$--model. The resulting $M$--$L_X$ relation is shown in Figure
\ref{fi:lm}. From the one hand, this relation demonstrates that a well
defined relation between X--ray luminosity and mass indeed exist,
although with some scatter, thus confirming that $L_X$ can indeed be
used as a proxy of the cluster mass. From the other hand, the slope of
the relation is found to be steeper than the self--similar scaling,
thus consistent with the observed $L_X$--$T$ relation. 

\subsubsection{Using the optical luminosity}

The classical definition of optical richness of clusters is known to
be a poor tracer of the cluster mass (e.g.,
\cite{2001Natur.409...39B}). However, the increasing quality of
photometric data for the cluster galaxy population and the ever
improving capability of removing fore/background galaxies thanks to
larger spectroscopic galaxy samples have recently allowed different
authors to demonstrate the optical/near--IR luminosities to be as
reliable tracers of the cluster mass as the X--ray luminosity.

\begin{figure}
\centerline{
\hbox{
\psfig{file=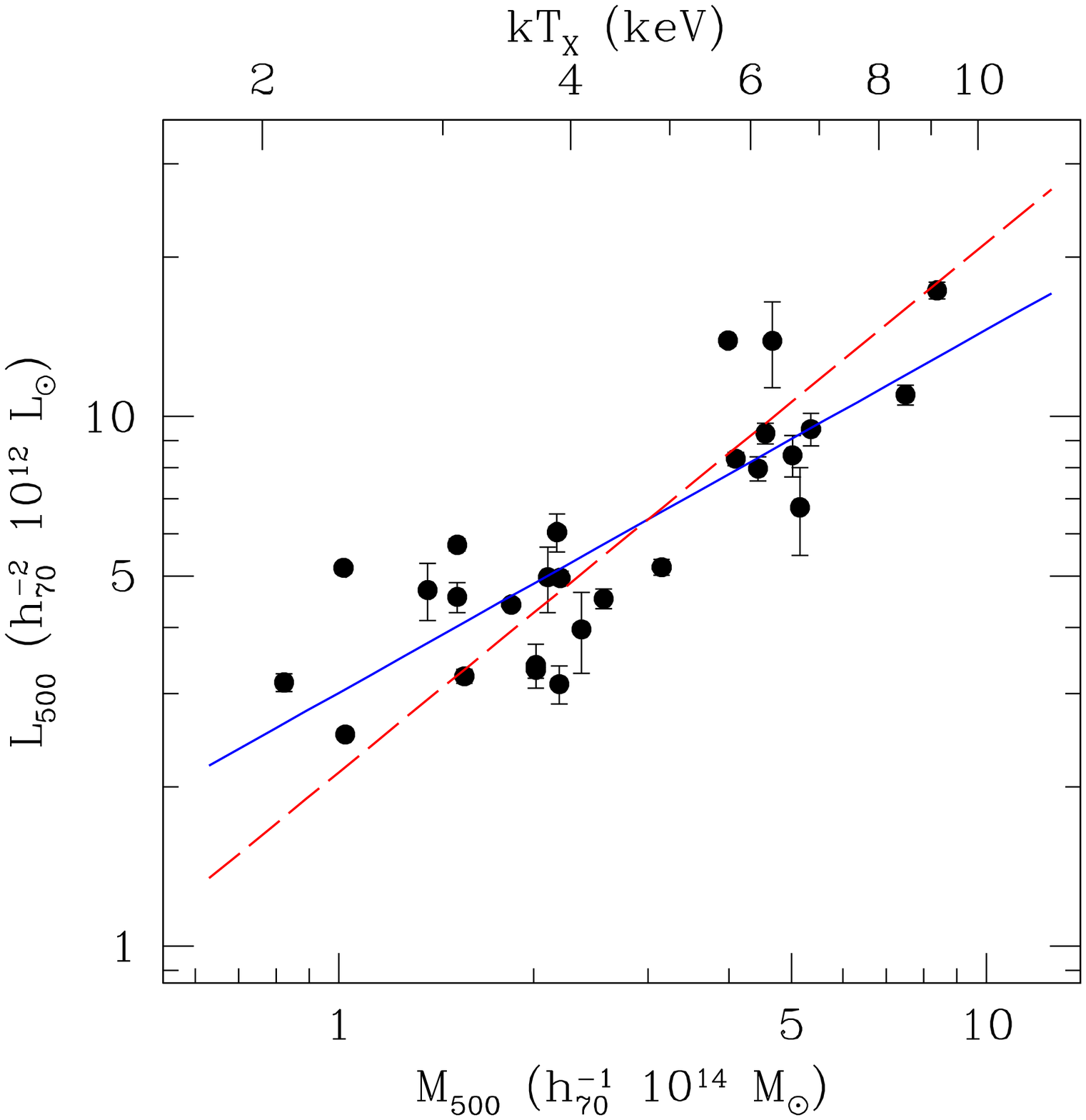,width=6.cm}
\psfig{file=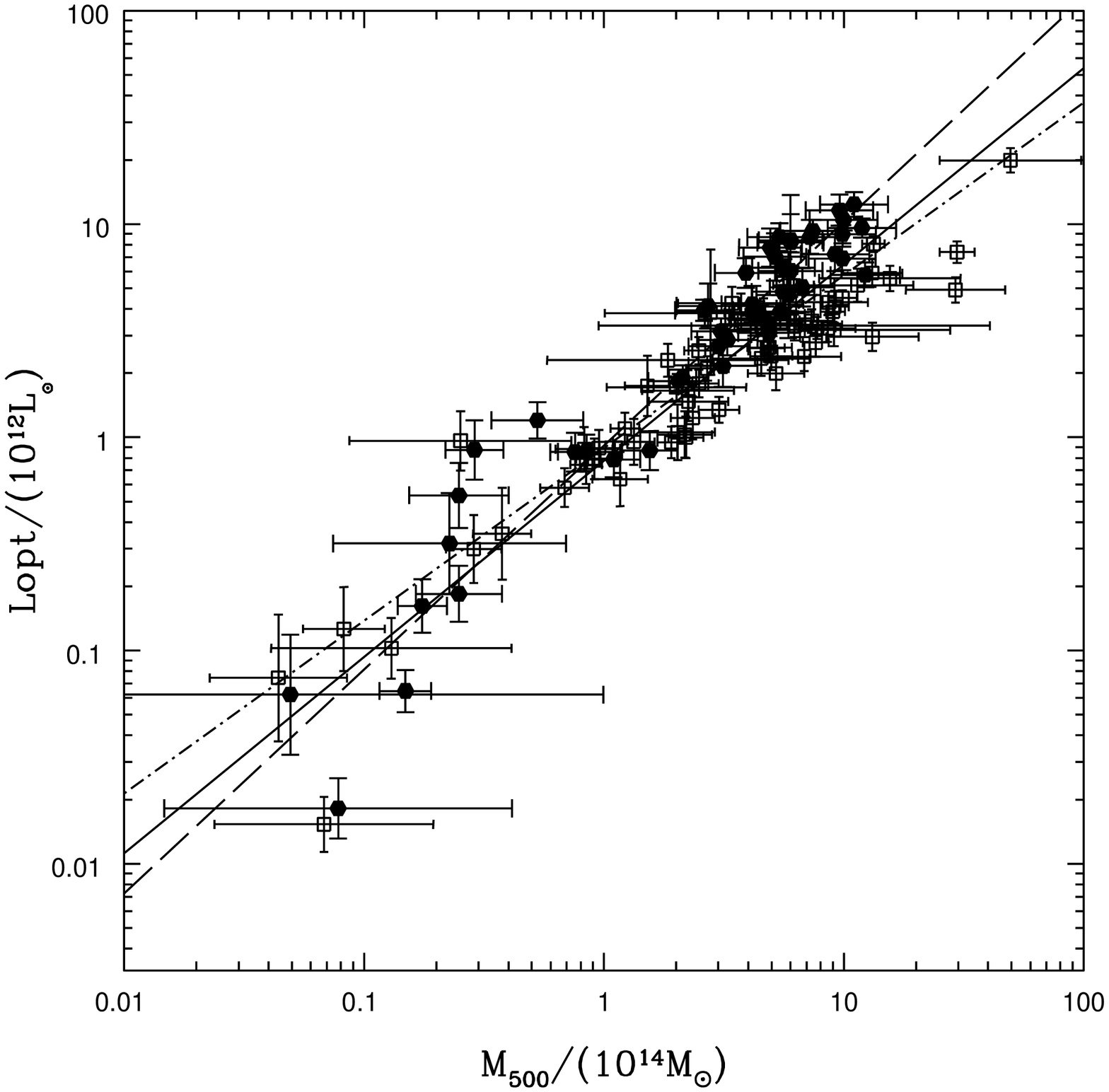,width=6.cm} 
}}
\caption{The relation between cluster masses and optical/near-IR
  luminosities. Left panel: $M_{500}$--$L_{500}$ relation from
  \cite{2003ApJ...591..749L}, using K--band data from the 2MASS survey
  and masses from X--ray data. The solid line is the best--fit power
  law, while the dashed line marks the unity slope. Right panel: the
  same relation, but in the $i$Sloan band. Open and filled points
  corresponds to mass estimates based on the SDSS spectroscopic data
  and on the $M$--$T_X$ relation, respectively.}
\label{fi:ml_opt} 
\end{figure}

Two examples of recent calibrations between optical/near--IR
luminosity and mass are shown in Figure \ref{fi:ml_opt}. In the left
panel we report the result presented in \cite{2003ApJ...591..749L},
based on K--band luminosites from the 2MASS and masses obtained by
applying the $M$--$T$ relation by \cite{2001A&A...368..749F}. Although
the data points are rather scattered, they define a clear
correlation. Quite interestingly, the best--fitting relation has a
slope shallower than unity, thus indicating the K--band mass-to-light
ratio is a (slightly) increasing function of the cluster mass. This
result is in line with previous results using optical luminosities
(e.g., \cite{2002ApJ...569..720G}) who found an increasing $M/L$ when
passing from galaxy groups to clusters of increasing richness.

Popesso et al. \cite{2005A&A...433..431P} analysed SDSS data for a 
set of clusters which
have been identified in the RASS. Their mass estimates come from both
X--ray temperature \cite{RE02.1} and from the velocity
dispersions as estimated from the SDSS spectroscopic data. The results
of their analysis for the $i$ band are shown in the right panel of
Fig. \ref{fi:ml_opt}. Again, the optical luminosity correlates quite
tightly with the cluster mass, with an intrinsic scatter which is
comparable to, or even smaller than that of the correlation between
X--ray luminosity and mass. 

These results highlight how cluster samples with precisely measured
optical luminosities can in principle be usefully employed to
constrain cosmological parameters. However, while X-ray luminosity
provides at the same time a tracer of cluster mass and a criterion to
precisely determine the sample selection function, the latter
quantity can be extracted from an optically selected sample only in a
rather indirect way.

\section{Constraints on cosmological parameters}
In this section we will review critically results on cosmological
constraints derived from different ways of tracing the cosmological
mass function of galaxy clusters. 

\subsection{The distribution of velocity dispersions}
A first determination of the mass function from velocity dispersions,
$\sigma_v$, of member galaxies has been attempted in
\cite{1993ApJ...411L..13B}. Girardi et al. \cite{1998ApJ...506...45G}
used a much larger sample of nearby clusters with measured velocity
dispersions to compare the resulting mass function with predictions
from cosmological models. The resulting relation between $\sigma_8$
and $\Omega_m$ was such that $\sigma_8\simeq 1$ for a fiducial value
of the density parameter $\Omega_m=0.3$. More recently, data for
nearby clusters, identified in the SDSS, have been used to calibrate a
relation between richness and velocity dispersion
\cite{2003ApJ...585..182B}. They compared the resulting
$\sigma_v$--distribution to the prediction of cosmological models and
found a significantly lower normalization of the power spectrum,
$\sigma_8\simeq 0.7$ for $\Omega_m=0.3$. Such differences from
different analyses highlight the presence of systematic uncertainties
in the relation between mass and observables (i.e., velocity
dispersion and richness).

The application of this method to distant clusters has been applied so
far only to the CNOC sample \cite{1997ApJ...479L..19C}, which
comprises 17 clusters selected from the EMSS out to $z\simeq
0.6$. Still to date, this is the only sample of distant clusters, with
calibrated selection function, for which velocity dispersions have
been reliably measured. Bahcall et al. \cite{1997ApJ...485L..53B}
pointed out that the resulting evolution of the mass function is
consistent with a low--density Universe. Borgani et
al. \cite{1999ApJ...527..561B} reanalysed this same sample and
emphasised that the uncertainties in the local normalization of the
mass function are large enough to make any constraints on $\Omega_m$
not significant.

\subsection{The temperature function}
The X-ray Temperature Function (XTF) is defined as the number density
of clusters with given temperature, $n(T)$. As long as a one-to-one
relation exist between temperature and mass, the XTF can be
related to the mass function, $n(M)$, by the relation 
\be
n(T)\,=\,n[M(T)]\,{dM\over dT}\,.
\label{eq:xtf}
\ee
In this equation, the ratio $dM/dT$ is provided by the
relation between ICM temperature and cluster mass.

Measurements of cluster temperatures for flux-limited samples of
nearby clusters were first presented in \cite{1991ApJ...372..410H}. These
results have been subsequently refined and extended to larger samples
with the advent of {\it ROSAT}, {\it Beppo--SAX} and, especially, {\it
  ASCA}.  XTFs have been computed for both nearby 
(e.g.,
\cite{1998ApJ...504...27M,2001MNRAS.325...77P,2003MNRAS.342..163P,2002A&A...383..773I})
and distant (e.g., \cite{1998MNRAS.298.1145E,1999ApJ...523L.137D,2000ApJ...534..565H,2004ApJ...609..603H})
clusters, and used to constrain cosmological models. The starting
point in the computation of the XTF is inevitably a flux-limited
sample for which the searching volume of each cluster can be computed.
Then the $L_X-T_X$ relation and its scatter is used to derive a
temperature limit from the sample flux limit.

Once the XTF is measured from observations, eq.(\ref{eq:xtf}) is used
to infer the mass function and, therefore, to constrain cosmological
models. A slightly different but conceptually identical approach, has
been followed in \cite{RE02.1}, where masses for a flux--limited
sample of nearby bright RASS clusters have been computed by applying
the assumption of hydrostatic equilibrium, thereby expressing their
results directly in terms of mass function, rather than of XTF.

\begin{figure}
\hbox{
\psfig{figure=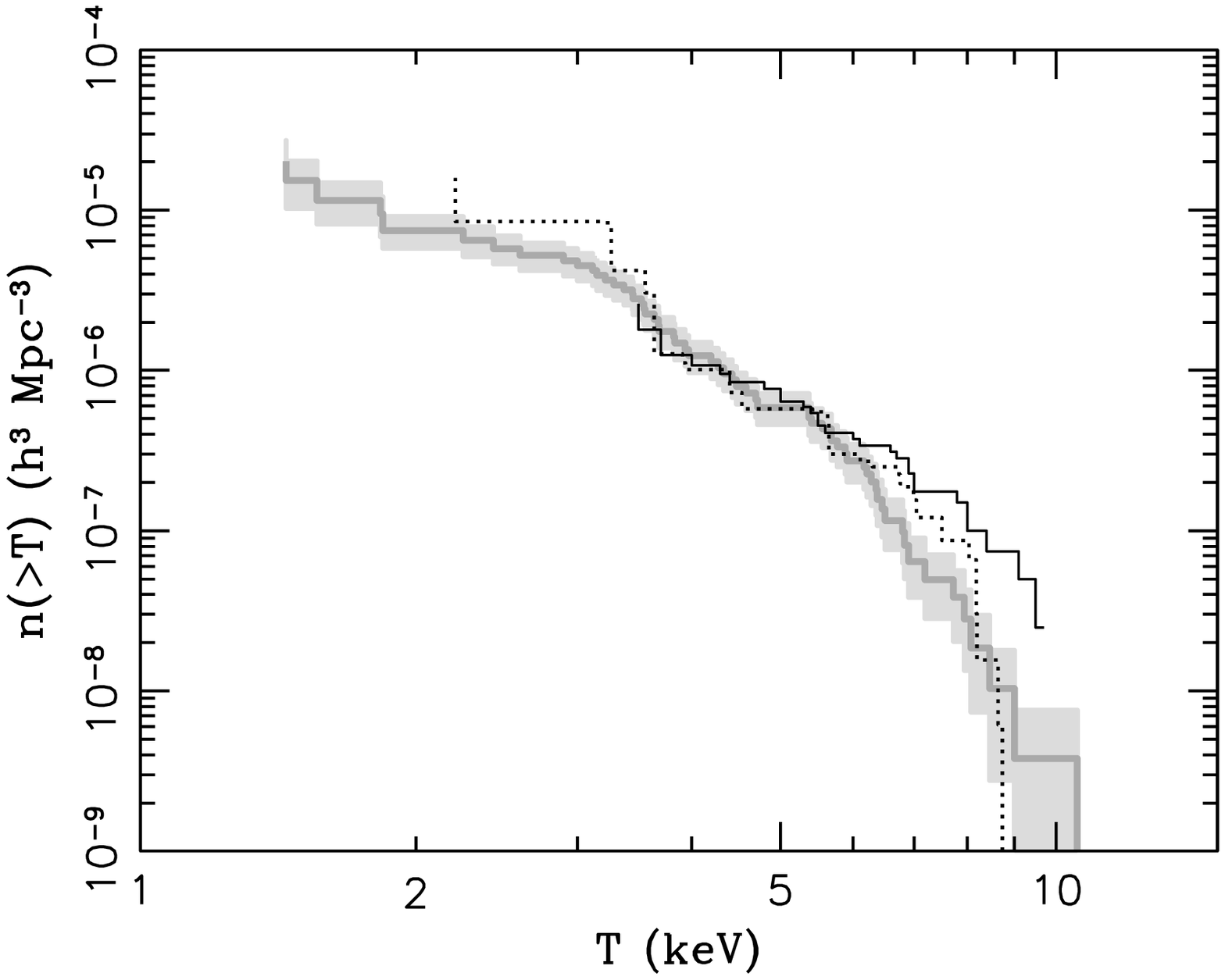,width=6.5truecm}
\psfig{figure=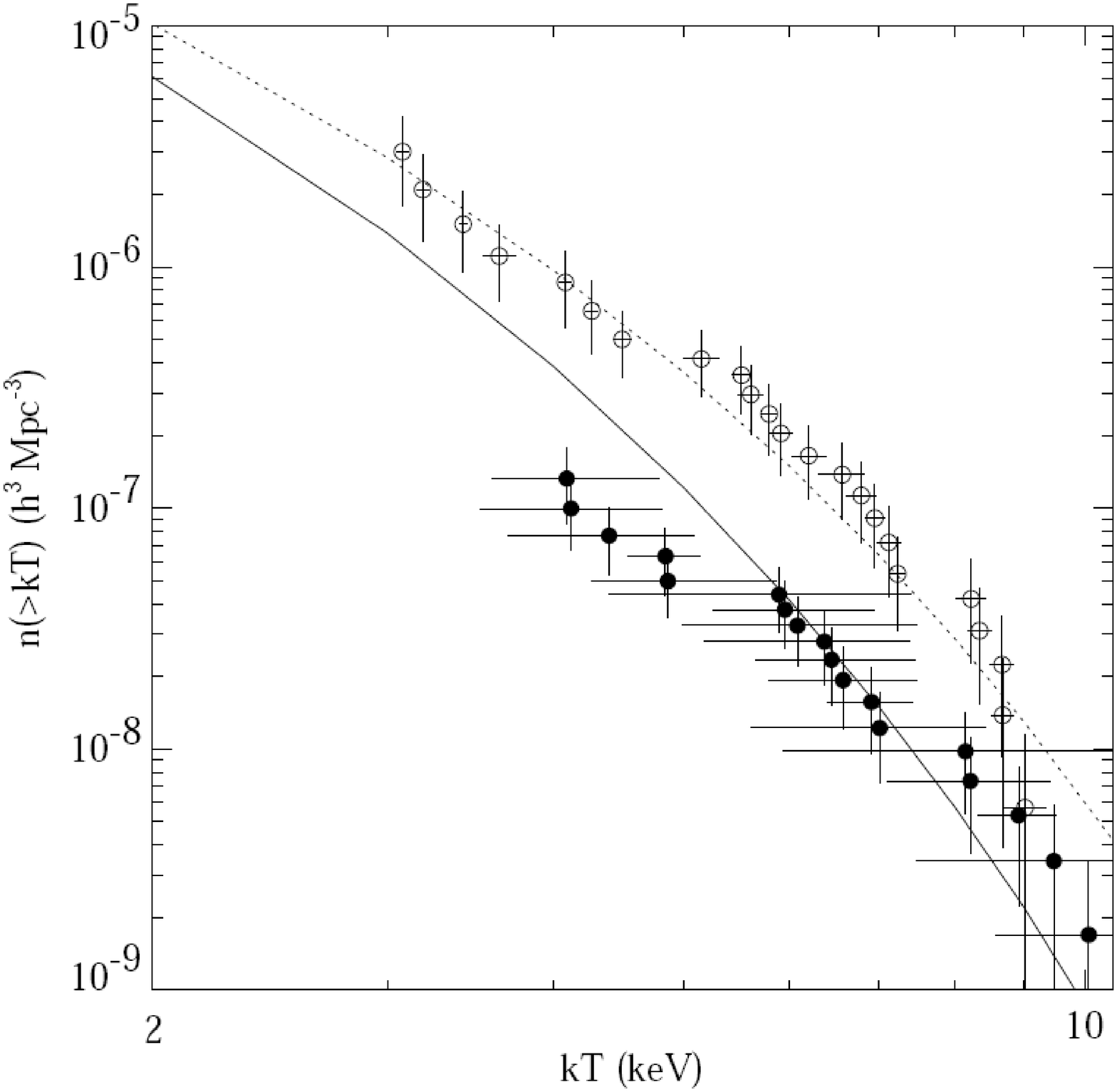,width=5.5truecm} 
}
\caption{Left panel: a comparison between the XTF for nearby clusters
  from \cite{2002A&A...383..773I} (shaded area), 
  \cite{1998ApJ...504...27M} (solid line)and 
  \cite{2000ApJ...534..565H} (dotted line). 
  Right panel: the evolution of the XTF
  from \cite{2004ApJ...609..603H}. Open and filled circles are for
  the local and the distant cluster sample, respectively. }
\label{fi:xtf}
\end{figure}

Oukbir and Blanchard \cite{1992A&A...262L..21O} first suggested to use
the evolution of the XTF as a way to constrain the value of
$\Omega_m$.  Several independent analyses converge now towards a mild
evolution of the XTF, which is interpreted as a case for a
low--density Universe, with $0.2\mincir \Omega_m \mincir 0.6$. An
example is reported in the right panel of Figure \ref{fi:xtf} (from
\cite{2004ApJ...609..603H}), which shows the comparison between the
XTFs of the sample of nearby clusters \cite{1991ApJ...372..410H} and a
sample of EMSS clusters with ASCA temperatures.

A limitation of the XTFs presented so far is the limited sample size
(with only a few $z\magcir 0.5$ measurements), as well as the lack of
a homogeneous sample selection for local and distant clusters. By
combining samples with different selection criteria one runs the risk
of altering the inferred evolutionary pattern of the cluster
population. This can even give results consistent with a
critical--density Universe
\cite{1997ApJ...488..566C,1999MNRAS.303..535V,2000A&A...362..809B}.

Besides the determination of the matter density parameter, the
observational determination of the XTF also allows one to measure the
normalization of the power spectrum, $\sigma_8$. Assuming a fiducial
value of $\Omega_m=0.3$, different (sometimes discrepant)
determinations of $\sigma_8$ have been reported by different authors,
ranging from $\sigma_8\simeq 0.7$--0.8 (e.g.,
\cite{1998MNRAS.298.1145E,RE02.1,2004ApJ...609..603H}) to
$\sigma_8\simeq 1$ (e.g.,
\cite{1998ApJ...504...27M,2001MNRAS.325...77P}). Ikebe et
al. \cite{2002A&A...383..773I} compared different observational
determinations of the XTF for nearby clusters (see left panel of
Fig. \ref{fi:xtf}) and established that they all agree with each other
reasonably well. Although quite comfortable, this result highlights
that the discrepant results on the normalization of the
$\sigma_8$--$\Omega_m$ relation comes from the cosmological
interpretation of the observed XTF, and not from observational
uncertainties in its calibration. While the different model mass
functions (i.e., whether Press--Schechter, Jenkins et al. or
Sheth--Tormen) can in some cases account for part of the difference,
more in general the different results are interpreted in terms of the
different normalization of the $M$--$T$ relation to be used in
eq.(\ref{eq:xtf}) or to the way in which the intrinsic scatter and the
statistical uncertainties in this relation are included in the
analysis. We shall critically discuss these issues in Section
\ref{s:sys} below.

Substantially improved observational determinations of the XTF, and
correspondingly tighter cosmological constraints, are expected to
emerge with the accumulations of data on the ICM temperature from the
Chandra and XMM--Newton satellites. Thanks to the much improved
sensitivity of these X--ray telescopes with respect to ASCA,
temperature gradients can be measured for fairly large sets of nearby
and medium--distant ($z\mincir 0.4$) clusters, thus allowing more
precise determinations of cluster masses. At the same time, reliable
measurements of global temperatures are now emerging for clusters out
to the highest redshifts where they have been secured (e.g., 
\cite{2004AJ....127..230R}). At the time of writing, several years after
the advent of the new generation of X--ray telescopes, no
determinations of the XTF from Chandra and XMM--Newton data have been
presented, a situation that is expected to change quite soon.

\subsection{The luminosity function}
Another method to trace the evolution of the cluster number density is
based on the X--ray luminosity function (XLF), $\phi(L_X)$, which is
defined as the number density of galaxy clusters having a given X--ray
luminosity.  Similarly to eq.(\ref{eq:xtf}), the XLF can be related to
the cosmological mass function of collapsed halos as
\be
\phi(L_X)\,=\,n[M(L_X)]\,{dM\over dL_X}\,,
\label{eq:xlf}
\ee
where $M(L_X)$ provides the relation between the observable $L_X$ and
the cluster mass. The above relation needs to be suitably modified in case
an intrinsic scatter exists in the relation between mass and
temperature (see Section \ref{s:sys}, here below). 

A useful observational quantity, that is related to the XLF, is given
by the flux number--counts, $n(S)$, which is defined as the number of
clusters per steradian, having measured flux $S$:
\be
n(S)\,=\,\left({c\over H_0}\right)^3\int_0^\infty dz \,{r^2(z)\over
E(z)}\, n[M(S,z);z]\,{dM\over dS}
\label{eq:ns}
\ee
(e.g., \cite{1997ApJ...490..557K})
where $r(z)$ is the radial coordinate appearing in the
Friedmann--Robertson--Walker metric:
\ba
r(z) & \!=\! & \int_0^z dz \,E^{-1}(z) ~~~~~;~~~~~\Omega_\Lambda=1-\Omega_m \nonumber \\
r(z) & \!=\! & {2\left[\Omega_mz+(2-\Omega_m)\,(1-\sqrt{1+\Omega_mz})\right]\over
\Omega_m^2(1+z) } ~~~;~~~\Omega_\Lambda=0\,.
\label{eq:rz}
\ea
The flux $S$ is related to the luminosity according to  
\be
S\,=\,{L_X\over 4\pi d_L^2(z)}\,,
\ee
where $d_L(z)=r(z)(1+z)$ is the luminosity distance at redshift $z$.

This quantity can be measured for a flux--limited samples without
having information on cluster redshift and provides useful
cosmological information in the absence of any spectroscopic optical
follow--up. A comparison between different observational
determinations of the flux number counts for both nearby and distant
cluster samples (e.g., \cite{2002ARA&A..40..539R}) show indeed a
quite good agreement.

Another quantity, which has been used to derive cosmological
constraints from flux--limited surveys, is the redshift distribution,
$n(z)$, which is defined as the number of clusters found in a survey
at a given redshift $z$:
\be
n(z) = \left({c\over H_0}\right)^3\,{r^2(z)\over
  E(z)}\,\int_{S_{lim}}^\infty dS\,f_{sky}(S)\,n[M(S,z);z]\,{dM\over
  dS}\,.
\label{eq:nz}
\ee 
In the above expression $S_{lim}$ is the limiting completeness
flux of the survey, while $f_{sky}(S)$ is the effective
flux--dependent sky--coverage appropriate for the considered
survey. Convolving the mass function with the sky coverage inside the
integral in the above equation is essential to properly account for
the different effective area covered at different fluxes, an aspect
which is apparently overlooked in some analyses 
(e.g., \cite{2003A&A...412L..37V}). 

In general, the advantage of using X-ray luminosity as a tracer of the
mass is that $L_X$ is measured for a much larger number of clusters in
samples with well-defined selection properties. As discussed in
Section~\ref{par:obs}, the most recent flux--limited cluster samples
contain now a fairly large ($\sim\!  100$) number of objects, which
are homogeneously identified over a broad redshift baseline, out to
$z\simeq 1.3$. This allows nearby and distant clusters to be compared
within the same sample, i.e. with a single selection function.
However, since the X--ray emissivity depends on the square of the gas
density, the relation between $L_X$ and $M_{\rm vir}$, which is based
on additional physical assumptions, is more uncertain than the $M_{\rm
vir}$--$\sigma_v$ or the $M_{\rm vir}$--$T$ relations.

\begin{figure}
\centerline{
\psfig{figure=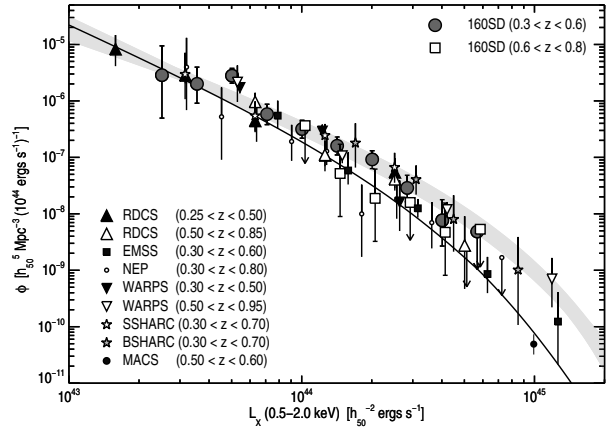,width=8.5truecm}
}
\caption{A compilation of XLF within different redshift intervals for
 independent X--ray flux--limited surveys (from
 \cite{2004ApJ...607..175M}). The shaded area shows the range of
 determinations of the local XLF, while the solid curve is the
 best--fitting evolving XLF from \cite{2004ApJ...607..175M}. }
\label{fi:xlf_evol}
\end{figure}

A useful parametrization for the relation between temperature and
bolometric luminosity can be casted in the form
\be
L_{bol}\, = \, L_6\,\left({T_X\over 6 {\rm keV}}\right )^\alpha(1+z)^A
\left({d_L(z)\over d_{L,EdS}(z)}\right)^2
\,10^{44} h^{-2}\lum \,,
\label{eq:lt}
\ee
with $L_6$ defining the normalization of the relation and $d_L(z)$ the
luminosity--distance at redshift $z$ for a given cosmology.

Analyses of the number counts from different X-ray flux--limited
cluster surveys showed that the resulting constraints on $\Omega_m$
are rather sensitive to the evolution of the mass--luminosity relation
\cite{1997ApJ...490..557K,1998MNRAS.295..769M,1999ApJ...527..561B}. On
the other hand, other authors
\cite{1998A&A...329...21S,1999ApJ...518..521R,2003A&A...412L..37V}
analysed different flux--limited surveys and found results consistent
with $\Omega_m=1$.  Quite intriguingly, this conclusion is common to
analyses which combine a normalization of the local mass function,
using nearby clusters, and to the evolution of the mass function using
deep surveys. Clearly, any uncertainty in the calibration of the
selection functions when combining different surveys may induce a
spurious signal of evolution of the cluster population, possibly
misinterpreted as an indication for high $\Omega_m$.

\begin{figure}
\centerline{\psfig{figure=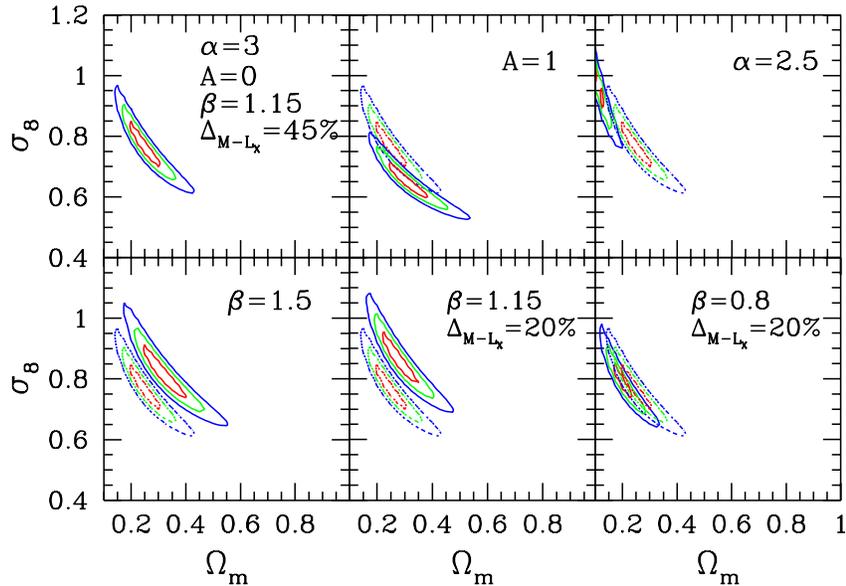,width=4.6in}}
\caption{Probability contours in the $\sigma_8$--$\Omega_m$ plane from
  the evolution of the X-ray luminosity distribution of RDCS
  clusters. The shape of the power spectrum is fixed to $\Gamma=0.2$
  \cite{2002ARA&A..40..539R}. Different panels refer to different
  ways of changing the relation between cluster virial mass, $M$, and
  X-ray luminosity, $L_X$, within theoretical and observational
  uncertainties (see also \cite{2001ApJ...561...13B}). The upper
  left panel shows the analysis corresponding to the choice of a
  reference parameter set.  In each panel, we indicate the parameters
  which are varied, with the dotted contours always showing the
  reference analysis. }
\label{fi:like}
\end{figure}

In order to overcome this potential problem, Borgani et
al. \cite{2001ApJ...561...13B} analyzed the RDCS sample to trace the
cluster evolution over the entire redshift range, $0.05\mincir
z\mincir 1.3$, probed by this survey, without resorting to any
external normalization from a different survey of nearby clusters
\cite{1999ApJ...517...40B}. They found $0.1\mincir \Omega_m\mincir
0.6$ at the $3\sigma$ confidence level, by allowing the $M$--$L_X$
relation to change within both the observational and the theoretical
uncertainties. In Figure \ref{fi:like} we show the resulting
constraints on the $\sigma_8$--$\Omega_m$ plane (from
\cite{2002ARA&A..40..539R}) and how they vary by changing the
parameters defining the $M$--$L_X$ relation: the slope $\alpha$ and
the evolution $A$ of the $L_X$--$T$ relation (see Equation
\ref{eq:lt}), the normalization $\beta$ of the $M$--$T$ relation (see
Equation \ref{eq:mteke}), and the overall scatter
$\Delta_{M-L_X}$. Flat geometry is assumed here, i.e.
$\Omega_m+\Omega_\Lambda=1$.

Similar results have been obtained by
combining information on clustering properties and the
redshift distribution from the the REFLEX cluster survey
\cite{2002MNRAS.335..807S}, thus providing
$\sigma_8\simeq 0.7$ and $\Omega_m\simeq 0.35$. One should however
notice that, since these constraints are derived from nearby clusters,
the corresponding estimate of $\Omega_m$ comes from the shape of the
CDM power spectrum, rather than from the growth rate of
perturbations. It is rather reassuring that dynamical and geometrical
constraints on $\Omega_m$ are in fact consistent with each other.

Constraints on $\Omega_m$ from the cluster X--ray luminosity and
temperature distribution are thus in line with the completely
independent constraints derived from the baryon fraction in clusters,
$f_{\rm bar}$ (e.g.,
\cite{1993Natur.366..429W,2003A&A...398..879E,2004MNRAS.353..457A}).

\subsection{The gas mass function}
An alternative way of tracing the mass function of galaxy clusters is
based on using as its proxy the mass function of the cluster gas
content \cite{2003ApJ...590...15V}. This method is based on the
assumption that galaxy clusters are fair containers of cosmic
baryons. Similarly to the method based on the baryon fraction, it
relies on the knowledge of the cosmic baryon fraction, either provided
by data on the deuterium abundance in high--redshift absorption
systems (e.g., \cite{2003ApJS..149....1K}) combined with predictions
of primordial nucleosynthesis, or from the spectrum of CMB
anisotropies (e.g.,
\cite{2005astro.ph..7503M,2006astro.ph..3449S}). This method has the
potential advantage that cluster gas mass is an easier quantity to
measure than the total collapsed mass, since it is essentially related
to the total cluster emissivity.

If we define $n_b(M_b)$ to be the baryonic mass function and $n(M)$
the total mass function, by definition we have $n_b(M_b)=n(\Omega_m
M_b/\Omega_b)$. Therefore, once $n_b(M_b)$ and $\Omega_b$ are known
from observations, the total mass function can be computed as a
function of $\Omega_m$, thereby treated as a fitting parameter.

While this method has the remarkable advantage of avoiding the
uncertainties related to direct estimates of the total collapsed mass,
it is affected by possible violations of the assumption of
universality of the baryon content of clusters. Indeed, while this
assumption should be valid for suitably relaxed and massive clusters, it
may be less so when considering all objects belonging to a
flux--limited sample, thus including also relatively small clusters
and structures with a complex non--relaxed dynamics. 

This method was applied \cite{2004ApJ...601..610V} to a set of bright
clusters selected from the RASS \cite{RE02.1} and found
$\sigma_8=0.72\pm 0.04$ with $\Omega_mh=0.13 \pm 0.07$. Chandra
observations were included in the analysis for a set of clusters
extracted from the 160 deg$^2$ survey \cite{2003ApJ...590...15V}.
They found an evolution of the gas mass function, which is consistent
with a flat cosmological model with $\Omega_m=0.3$.

We emphasize here that the above different methods used to reconstruct
the mass function of galaxy clusters consistently prefer relatively
low values of $\sigma_8$, in the range 0.7--0.8. Quite remarkably,
such values have been shown now to be required by the 3--years WMAP
data release (\cite{2006astro.ph..3449S}).

\subsection{Including uncertainties in the analysis}
\label{s:sys}

As we have discussed in the previous sections, most of the analyses of
the cluster populations converge toward a low--density model, with
$\Omega_m\sim0.3$. However, significant differences exist between
different determinations of the normalization of the power spectrum,
$\sigma_8$, which amount to up to $\sim 20$ per cent. These
differences are much larger that the statistical uncertainties
associated to the finite number of clusters included in the samples,
thus indicating that they arise from unaccounted sources of error,
which affects the analyses.

\begin{figure}
\hbox{
\psfig{figure=Figs/sigom_ikebe.eps,angle=270,width=6.cm}
\psfig{figure=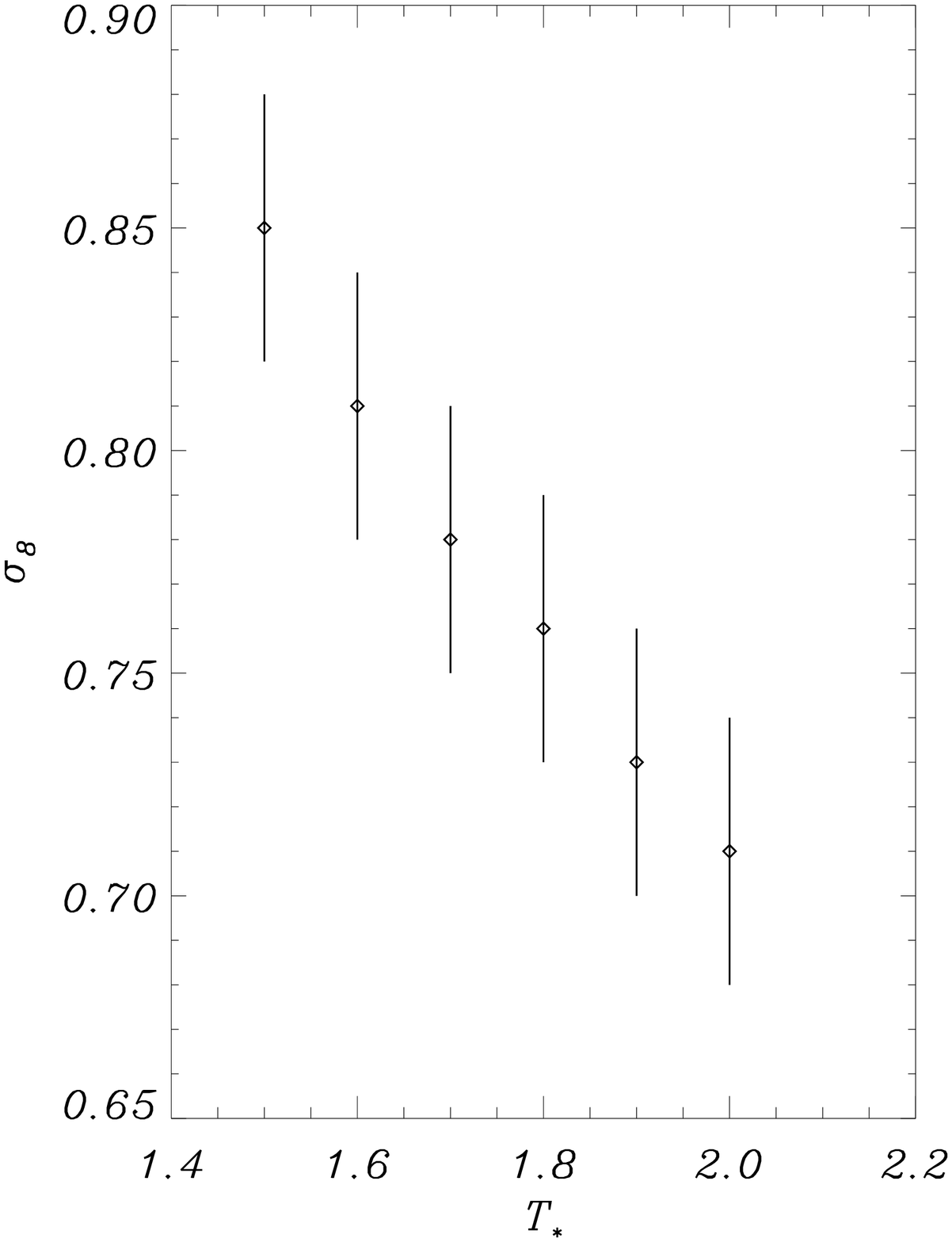,height=7.cm} 
     }
     \caption{Left panel: the dependence of the best--fitting values
       of $\Omega_m$ and $\sigma_8$, from the XTF of
       \cite{2002A&A...383..773I}, upon different determinations of
       the mass--temperature relation, from both observational data
       and from hydrodynamical simulations of galaxy clusters. Right
       panel: the dependence of $\sigma_8$ on the normalization of the
       mass--temperature relation, in the XTF analysis by
       \cite{2003MNRAS.342..163P}.}
\label{fi:s8comp}
\end{figure}

For instance, the role of the uncertain normalization of the
mass--temperature relation in the determination of $\sigma_8$ (at
fixed $\Omega_m$) from the XTF analysis has been emphasized by
different authors (e.g.,
\cite{2002A&A...383..773I,2003MNRAS.342..163P,2002MNRAS.337..769S}). Increasing
the normalization of the $M$--$T$ relation implies that a larger mass
corresponds to a fixed temperature value. As a consequence, an
observed XTF translates into a larger mass function, therefore
implying a larger $\sigma_8$ (at fixed $\Omega_m$). Since results from
hydrodynamical simulations generally imply a larger $M$--$T$
normalization, a larger $\sigma_8$ is expected when using in the
analysis the simulation predictions. The left panel of Figure
\ref{fi:s8comp} (from \cite{2002A&A...383..773I}) shows how the best
fitting values of $\sigma_8$ and $\Omega_m$ from the local XTF change
as one uses different mass--temperature relations, taken from both
observations and simulations. The right panel of Fig.\ref{fi:s8comp}
(from \cite{2003MNRAS.342..163P}) show the dependence of $\sigma_8$ on
the normalization of the mass--temperature relation. Note that this
normalization is allowed here to vary over the range encompassed by
observational and simulation results. The range of variation of
$\sigma_8$ induced by the uncertainty in the $M$--$T$ relation is at
least comparable to the purely statistical uncertainty, as indicated
by the errorbars.

One may wonder why not relying only on the observational determination
of the $M$--$T$ relation, instead of considering also simulation
results. As we shall discuss in Section \ref{s:fut}, observational
results can not necessarily provide the most reliable determination of
the $M$--$T$ scaling.

The effect of changing the normalization of the $M$--$T$ relation on
the analysis of flux--limited samples through the XLF evolution is
shown in the lower left panel of Fig.\ref{fi:like}. If this
normalization is reduced by $\sim\! 30\%$, the resulting $\sigma_8$
decreases by $\sim\!  20\%$.

It is clear that, any uncertainty, both statistical and systematic, in
the fitting parameters describing the scaling relations between mass
and observable, must be included in the analysis by marginalizing over
the probability distribution function of these parameters. Let ${\bf
  \Omega}$ be the set of cosmological parameters that we want to
constrain, and ${\bf W}$ the set of parameters which define a scaling
relation between mass $M$ and an observable $X$ (i.e., $\sigma_v$,
$L_X$ or $T$). Let us also call $P({\bf W})$ the prior distribution
for the uncertainties in the $M$--$X$ relation. If $\chi^2({\bf
  \Omega}, {\bf W})$ gives the goodness of fit provided by the choice
${\bf \Omega}$ of the cosmological parameters, for a given $M$--$X$
relation, then the goodness of fit after marginalizing over the
uncertainties in $M$--$X$ reads
\be
\chi^2({\bf \Omega})\,=\,{\int \chi^2({\bf \Omega}, {\bf W}) \,P({\bf
  W}) \,d{\bf W}\over \int P({\bf
  W}) \,d{\bf W}}\,.
\label{eq:marg}
\ee
Of course, the marginalization generally induces an increase of the
uncertainties in the cosmological parameters. Furthermore, one needs
to have a reliable modeling of both size and distribution of the
errors (i.e. whether they have a uniform, a Gaussian, or a more
peculiar distribution).

\begin{figure}
\hbox{\psfig{figure=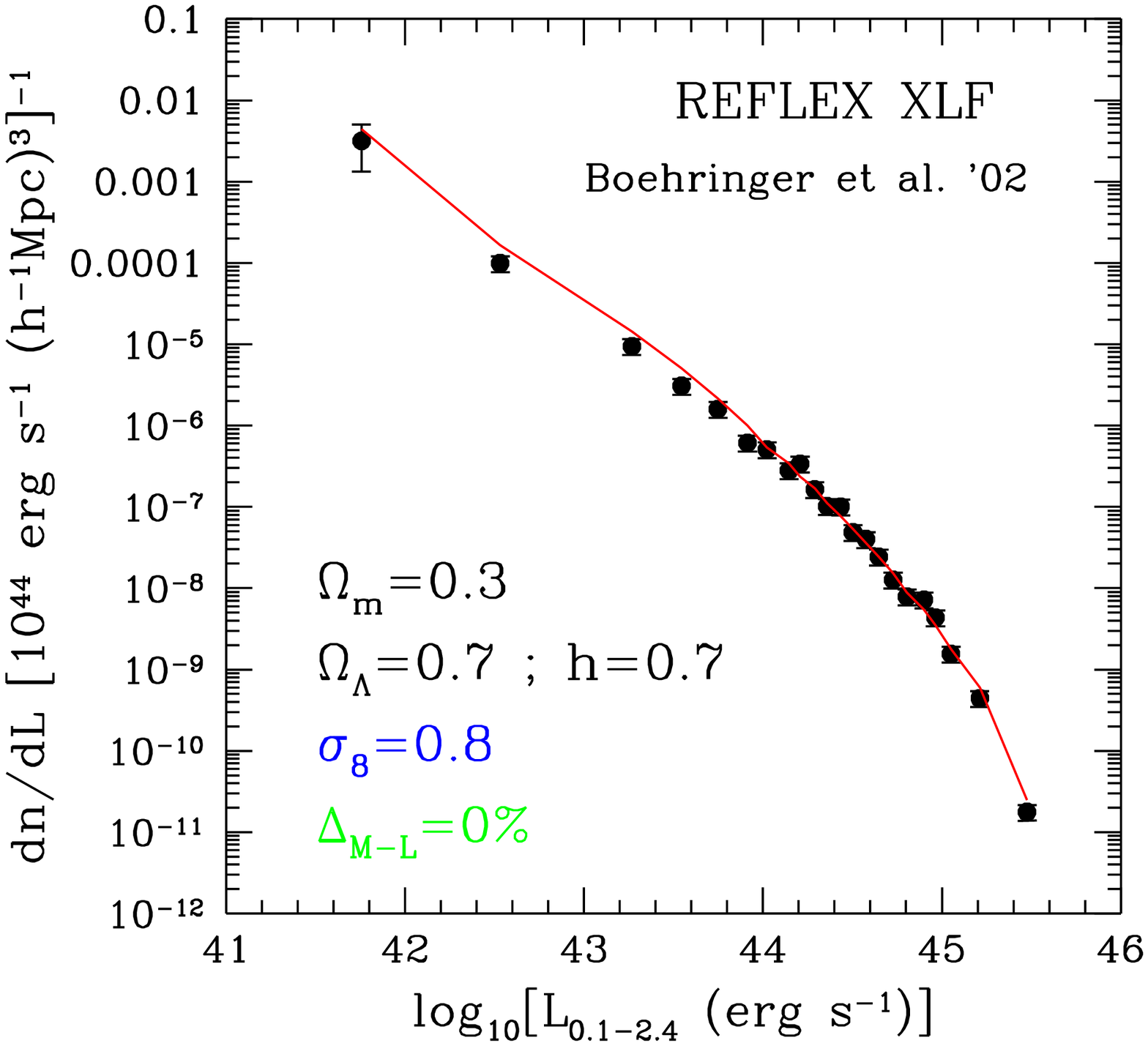,width=6.truecm}
      \psfig{figure=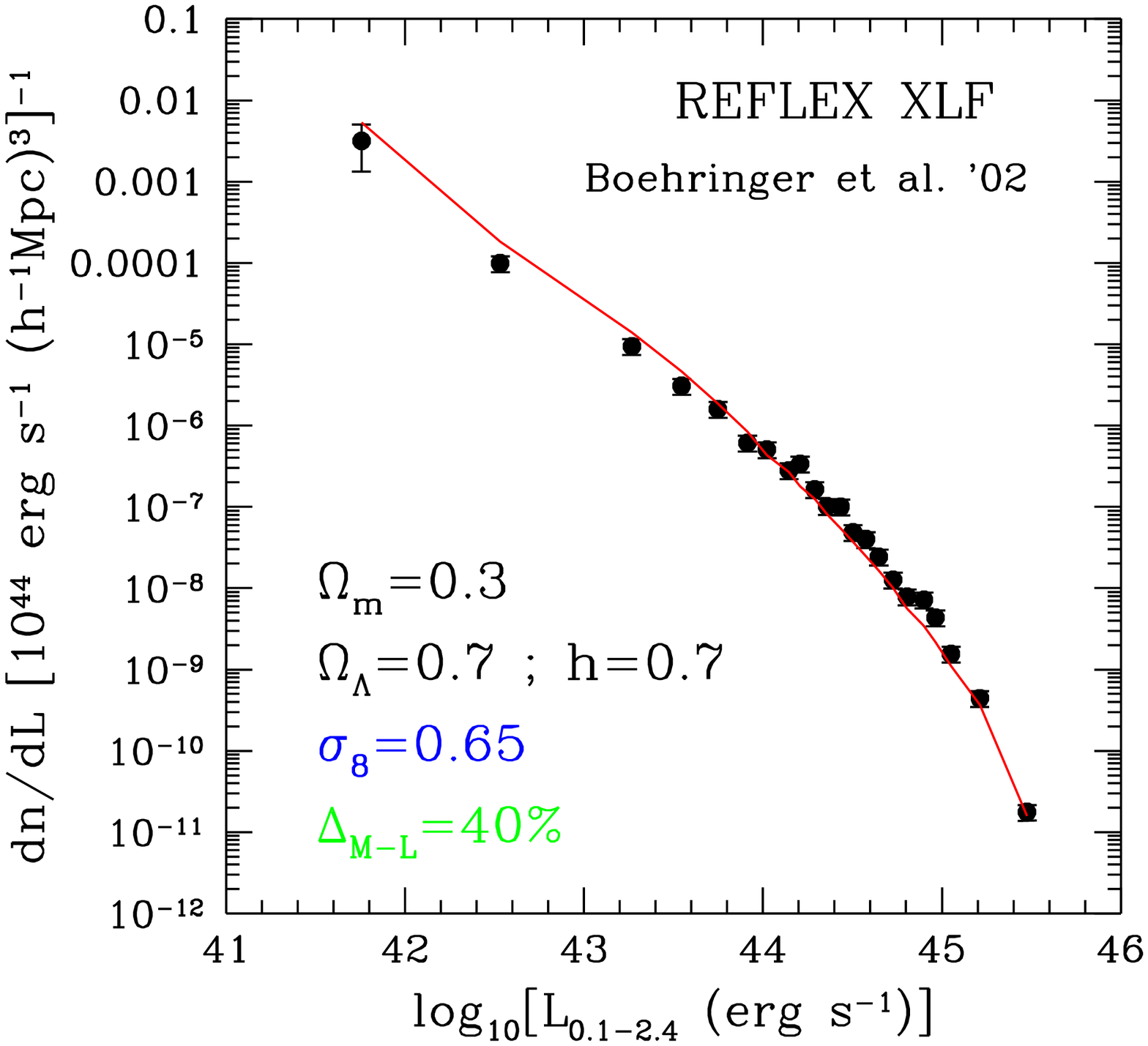,width=6.truecm} 
     }
     \caption{The dependence of $\sigma_8$ on the intrinsic scatter
       assumed in the relation between X--ray luminosity and mass. The
       two panels show the comparison between model predictions and
       the observed XLF from the REFLEX sample
       \cite{2002ApJ...566...93B}. The left panel assumes vanishing
       scatter while the right panel assumes a 40 per cent intrinsic
       scatter. The best fitting values of $\sigma_8$ are reported in
       both cases.}
\label{fi:s8scat}
\end{figure}

Besides the errors in the parameters defining the scaling relations, a
different source of uncertainty is provided by the intrinsic scatter
in these relations. Intrinsic scatter has the effect of widening the
range of possible masses which correspond to a given value of the
observable quantity. This effect can be included in the analysis by
convolving the theoretical mass function with the distribution of the
scatter itself (e.g., \cite{2001ApJ...561...13B,2005PhRvD..72d3006L}).
Let $\Psi$ be an observable quantity and $\phi(\Psi)$ its distribution
(i.e., XLF or XTF), to be compared with observations. Also, let
$P(M_\Psi|M;z)$ be the probability of assigning a mass $M_\Psi$ to a
cluster of true mass $M$, at redshift $z$, from the observable $\Psi$,
for a given $M$--$\Psi$ relation. Therefore, the model prediction for
the distribution $\phi(\Psi)$, to be compared with its observational
determination, is given by the convolution of the cosmological mass
function with the distribution of the intrinsic scatter:
\be
\phi(\Psi)d\Psi\,=\,\int dM_\Psi n(M_\Psi,z)\,P(M_\Psi|M;z)\,{dM\over d\Psi}\,d\Psi
\label{eq:conv}
\ee
where $n(M,z)$ is the cosmological mass function at redshift $z$. 
If one makes the standard assumption of Gaussian scatter in the
log--log plane, then
\be
P(M_\Psi|M)\,=\,\left( 2\pi \sigma_{\ln M}^2\right)^{-1/2} \exp\left[
  -x^2(M_\Psi)\right] \,,
\label{eq:logsca}
\ee
where $x(M_\Psi)=(\ln M_\Psi -\ln M)/(\sqrt 2 \sigma_{\ln M})$ and
$\sigma_{\ln M}$ is the r.m.s. intrinsic scatter. The effect of this
convolution is that of increasing $\phi(\Psi)$, for a fixed $n(M)$, as
the scatter increases. Therefore, assuming a progressively larger
scatter in the $M$--$\Psi$ relation implies a progressively lower
$\sigma_8$ (at fixed $\Omega_m$). An illustrative example of the
effect of intrinsic scatter on the determination of $\sigma_8$ is
reported in Figure \ref{fi:s8scat}. In the left panel we show the
REFLEX XLF \cite{2002ApJ...566...93B} along with the prediction of
the best fitting cosmological model for a given choice of the
$M$--$L_X$ relation, after assuming vanishing intrinsic scatter in
this relation. In the left panel, we show the same comparison, but
assuming an Gaussian--distributed intrinsic scatter of 40 per cent in
the $M$--$L_X$ scaling. As expected, adding the scatter has the effect
of increasing the predicted luminosity function, so that $\sigma_8$
has to be lowered from 0.8 to 0.65 to recover the agreement with
observations.
This example highlights that a good calibration of the intrinsic
scatter in the scaling relations can be as important as determining
the best--fitting amplitude and slope of these relations.

\section{The future}
\label{s:fut}
A new era for cosmology with galaxy clusters is now starting.  High
sensitivity surveys for blind SZ identification over fairly large
contiguous area, $\sim 100$ deg$^2$, have have already started or are
planned in the coming years (see the lectures by M. Birkinshaw in this
volume). Also, the Planck satellite will survey the whole sky,
although at a much lower sensitivity, and provide a large set of
clusters identified through the SZ effect. These surveys promise to
identify several thousands clusters, with a fair number of objects
expected to be found at $z>1$.  In the optical/near-IR bands,
imaging with dedicated telescopes with large field of view will also
allow to secure a large number of distant clusters. At the same time,
X--ray observations over contiguous area (e.g.,
\cite{2005MNRAS.363..675W} and ``serendipitous'' searches from
XMM--Newton (e.g., \cite{2005ApJ...623L..85M}) and Chandra (e.g.,
\cite{2002A&A...396..397B}) archives will ultimately cover several
hundreds deg$^2$ down to flux limits fainter than those reached by the
deepest {\it ROSAT} pointings. Preliminary results suggest that
identification of $z>1$ clusters may eventually become routine
\cite{2005ApJ...623L..85M}. Ultimately, they will lead to the
identification of several thousands clusters.

Optimized optics for wide--field X--ray imaging have been originally
described in a far-reaching paper by Burrows et
al. \cite{1992ApJ...392..760B} and proposed for the first time to be
implemented in a dedicated satellite mission in the mid 90's. There is
no doubt that this would be the right time to plan a dedicated
wide--field X--ray telescope, which should survey the sky over an area
of several thousands deg$^2$, with a relatively good, XMM--like or
better, point spread function and a low background. This instrument
would be invaluable for studies of galaxy clusters, thanks to its
ability of both identifying extended sources with low surface
brightness. Several missions with a similar profile have been
proposed, although none has been approved so far. Still, the community
working on galaxy clusters is regularly proposing this idea of
satellite to different Space Agencies (e.g.,
\cite{2005astro.ph..7013H}).

The samples of galaxy clusters obtainable from large--area SZ and
X--ray surveys contain in principle so much information to allow one
to constrain not only $\Omega_m$ and $\sigma_8$, but also the Dark
Energy (DE) content of the Universe (see, e.g.,
\cite{2003RSPTA.361.2497S,2005LNP...653..141S} for introductory
reviews on Dark Energy). The equation of state of DE is written in the
form $p=w\rho$, where $p$ and $\rho$ are the pressure and density
terms, respectively. The parameter $w$ must take values in the range
$-1/3 > w \ge -1$ for the DE to provide an accelerated cosmic
expansion. Constraining the value of $w$ and its redshift evolution is
currently considered one of the most ambitious targets of modern
cosmology. Ones hope is to unveil the nature of the energy term which
dominates the overall dynamics of the Universe at the present time.

As we have mentioned, the limited statistics prevents current cluster
surveys to place significant constraints on $\Omega_\Lambda$.
Although this limitation will be overcome with future cluster surveys,
the question remains as to whether the systematic effects, discussed
in Section \ref{s:sys}, can be sufficiently understood. Different
lines of attack have been proposed in the literature, which should not
considered as alternative to each other.

\begin{figure}
\centerline{
\hbox{
\hspace{-2.3cm}
\psfig{figure=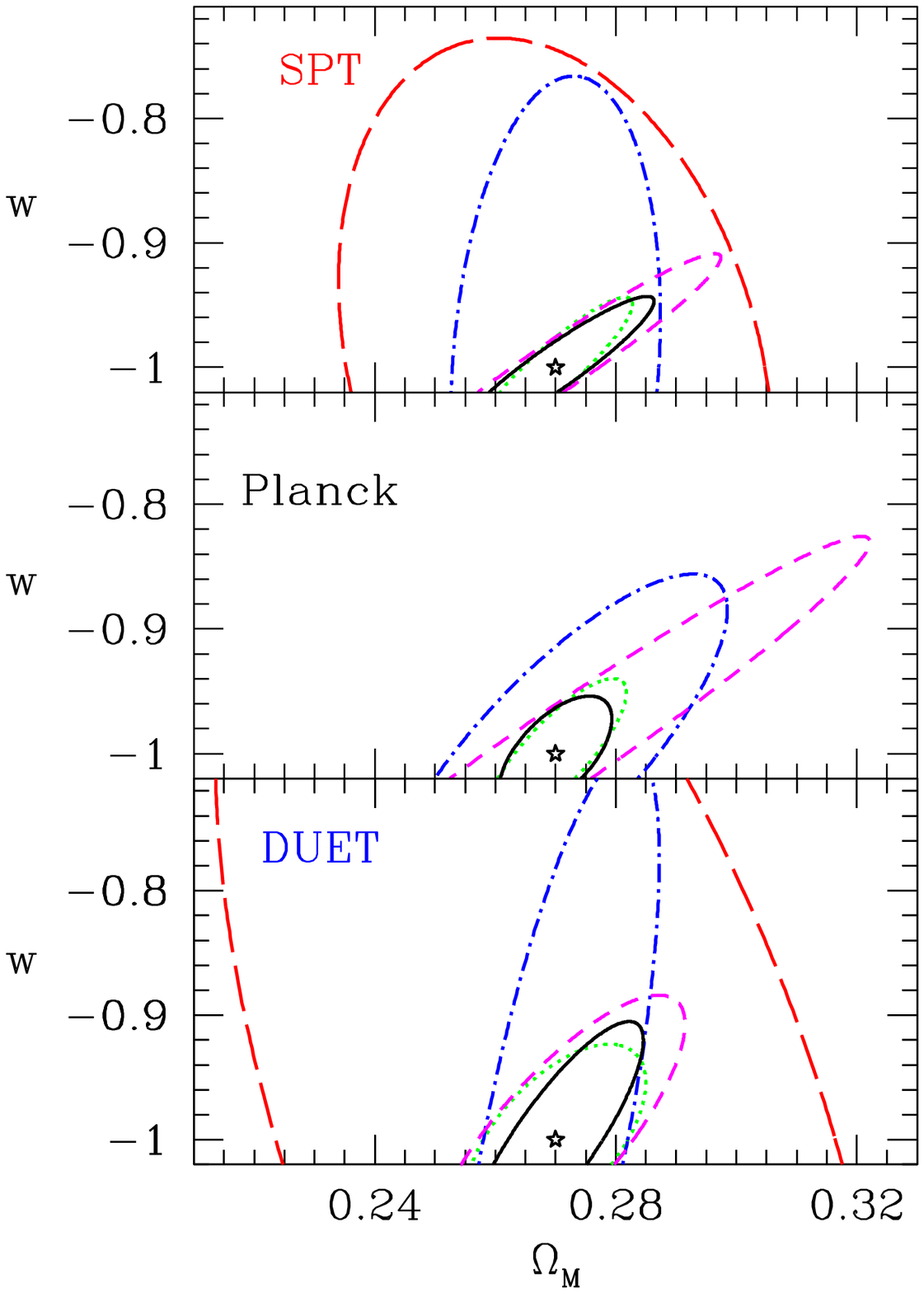,width=8.truecm}
\hspace{-2.3cm}
\psfig{figure=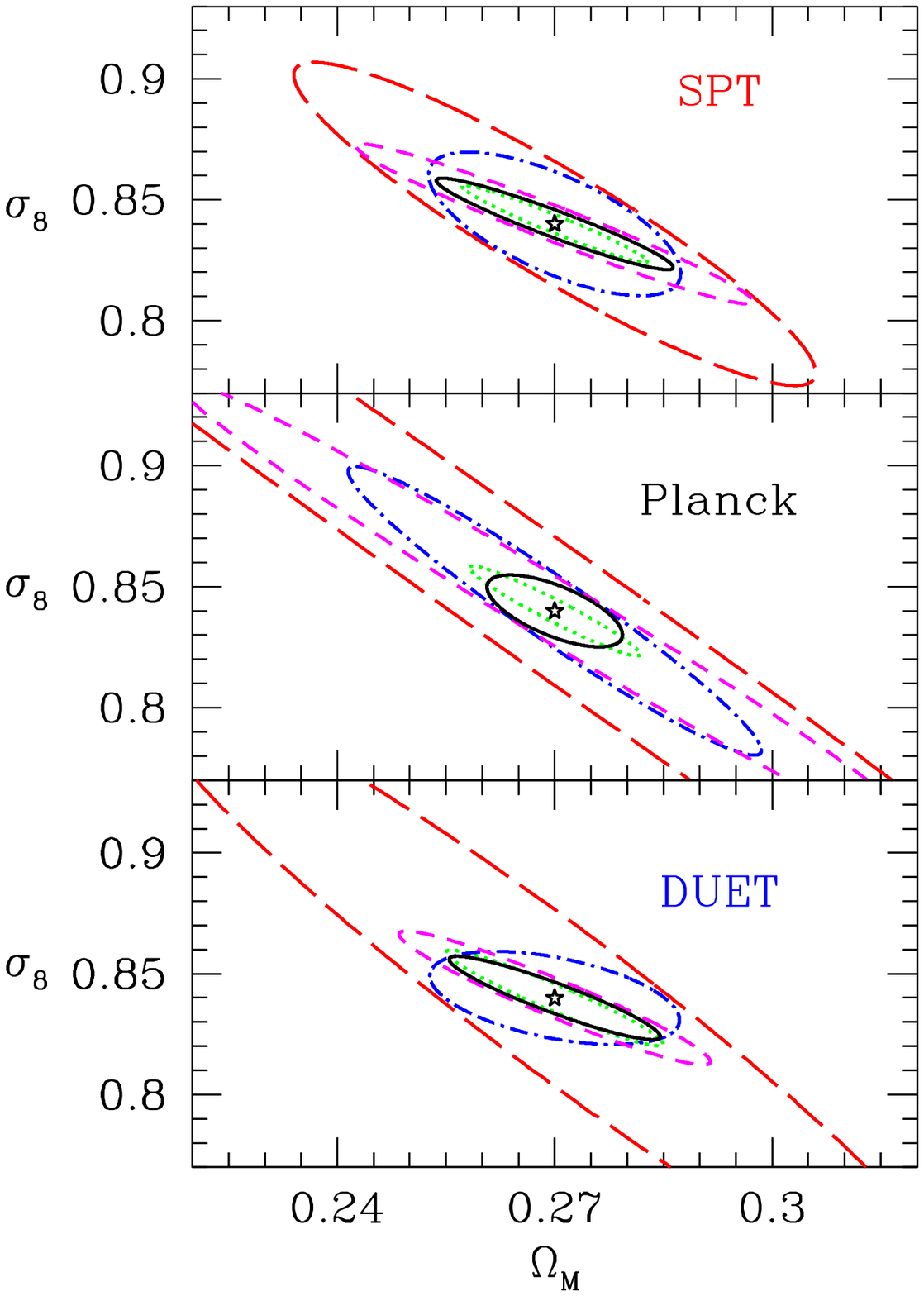,width=8.truecm}
}}
\caption{Constraints on cosmological parameters from the SZ SPT
  survey, from the SZ Planck survey and from a wide--area deep X--ray
  survey (from top to bottom panels; from
  \cite{2004ApJ...613...41M}). Dotted lines: no uncertainty in the
  relation between cluster mass and observables, using only the
  $dn/dz$ information; long--dashed line: using self--calibration on
  the $dn/dz$; dot--dashed line: as before, but also including the
  information on the cluster power spectrum; short--dashed line: using
  $dn/dz$ and a calibration of the mass--observable relation for 100
  clusters; solid line: all the information combined together.}
\label{fi:outocal}
\end{figure}

Majumdar and Mohr \cite{2004ApJ...613...41M} have proposed the
approach based on the so--called self--calibration (see also
\cite{2003ApJ...585..603M,2004PhRvD..70d3504L,2005PhRvD..72d3006L}). The
idea underlying this approach is that of parametrizing in a sensible
way the scaling relations between cluster observables and mass,
including the corresponding intrinsic scatter and its distribution. In
this way, the parameters describing these relations can be considered
as fitting parameter to be added to the cosmological parameters. As
long as the cluster samples are large enough, one should be able to
fit at the same time both cosmological parameters and those parameters
related to the physical properties of clusters. Figure
\ref{fi:outocal} (from \cite{2004ApJ...613...41M}) shows the
constraints that one can place on the $w$--$\Omega_m$ (left panel) and
on the $\sigma_8$--$\Omega_m$ planes, after marginalizing over the
other fitting parameters, from different SZ and X--ray surveys. In
each panel different contours indicate the constraints that one can
place by progressively adding information in the analysis. The main
message here is that combining information on the evolution of the
cluster population and on its clustering can place precision
constraints on cosmological parameters. These constraints can be
further tightened if follow up observations are available to precisely
measure masses for 100 clusters.

These forecasts nicely illustrates the potentiality of the
self--calibration approach for precision cosmology with future surveys
of galaxy clusters. Clearly, the robustness of these predictions is
inextricably linked to the possibility of accurately modeling the
relations between mass and observables.

A line of attack to this problem is based on using detailed
hydrodynamical simulations of galaxy clusters. The great advantages of
using simulations is that both cluster mass and observable quantities
can be exactly computed. Furthermore, the effect of observational
set--ups (e.g., response functions of detectors, etc.) can be included
in the analysis and their effect on the scaling relations
quantified. This approach has been applied by different groups in the
case of X--ray observations \cite{2004MNRAS.351..505G} to understand
the relation between the ICM temperature, as measured from the fitting
of the observed spectrum, and the ``true'' mass--weighted temperature
(e.g.,
\cite{2001ApJ...546..100M,2004MNRAS.354...10M,2005astro.ph..4098V}).
Furthermore, simulations can also be used to verify in detail the
validity of assumptions on which the mass estimators, applied to
observations, are based. The typical example is represented by the
assumption of hydrostatic equilibrium, discussed in Sect.\ref{s:he},
for which violations in simulations at the 10--20 per cent level have
been found (e.g.,
\cite{1996MNRAS.283..431B,2004MNRAS.351..237R,2004MNRAS.348.1078B,2004MNRAS.355.1091K}).

\begin{figure}[t]  
\begin{center}
\psfig{figure=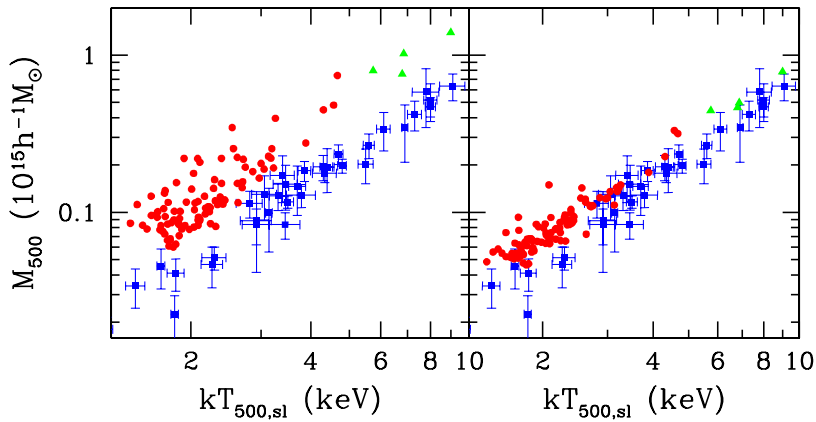,width=12.truecm}
\vspace{-5.5truecm}
\caption{The mass--temperature relation at $\bar\rho/\rho_{\rm
    cr}=500$, in simulations (filled circles and triangles) and for
    the observational data (squares with errorbars,
    \cite{2001A&A...368..749F}). The left panel is for the true masses
    of simulated clusters; the right panel is for masses of simulated
    clusters estimated by adopting the same procedure applied by
    Finoguenov et al. to observational data (from
    \cite{2005ApJ...618L...1R}).}
\label{fig:MvsT}
\end{center} 
\end{figure} 
   
An example of calibration of observational biases in the
mass--temperature relation, using hydrodynamical simulations, is shown
in Figure \ref{fig:MvsT} (from \cite{2005ApJ...618L...1R}), which
provides a comparison between the observed and the simulated $M$--$T$
relation. Simulations here include radiative cooling, star formation
and the effect of galactic winds powered by supernovae, and, as such,
provide a realistic description of the relevant physical
processes. The observational results, which are taken from
\cite{2001A&A...368..749F}, corresponds to mass estimates based on the
hydrostatic equilibrium for a polytropic $\beta$--model of the gas
distribution [eq.(\ref{eq:mhe})]. In both panels, the temperature has
been computed by using a proxy to the actual spectroscopic
temperature, the so--called spectroscopic--like temperature
\cite{2004MNRAS.354...10M}. The left panel shows the results when
exact masses of simulated clusters are used for the comparison. Based
on this result only, the conclusion would be that simulations do
indeed produce too high a $M$--$T$ relation, even in the presence of a
realistic description of gas physics. In the left panel, masses of
simulated clusters are computed instead by using the same procedure as
for observed clusters, i.e. by applying eq.(\ref{eq:betam}) for the
hydrostatic equilibrium of a polytropic $\beta$--model. Quite
remarkably, the effect of applying the observational mass estimator
has two effects. First, the overall normalization of the $M$--$T$
relation is decreased by the amount required to attain a reasonable
agreement with observations. Second, the scatter in the simulated
$M$--$T$ relation is substantially suppressed. This is the
consequence of the fact that eq.(\ref{eq:betam}) provides a one-to-one
correspondence between mass and temperature, while only the
cluster-by-cluster variations of $\beta$ and $\gamma$ account for the
intrinsic diversity of the cluster thermal structure.

This example illustrates how simulations can be usefully employed as
guidelines to study possible biases on observational mass
estimates. However, it is worth reminding that the reliability of
simulation results depends on our capability to correctly provide a
numerical description of all the relevant physical process. In this
sense, understanding in detail the (astro)physics of clusters is
mandatory in order to calibrate them as tools in the era of precision
cosmology.

As a concluding remark, we emphasize once more that a number of
independent analysesof the cluster mass function, which have been
realized so far, favor a relatively low normalization of the power
spectrum, with $\sigma_8\simeq 0.7$--0.8 for $\Omega_m\simeq 0.3$,
thus in agreement with the most recent WMAP results
(\cite{2006astro.ph..3449S}). This agreement must be considered as a
success for cluster cosmology and a strong encourgement for future
applications to large cluster surveys of the next generation.

\vspace{0.2truecm}
\noindent {\bf Acknowledgments.} I would like to thank the organizers
of the GH2005 School, David Hughes, Omar Lopez--Cruz and Manolis
Plionis for having provided an enjoyable and stimulating
environment. I also warmly thank Manolis Plionis and Piero Rosati for
a careful reading of the manuscript and useful suggestions to improve
it.

\printindex
\end{document}